\documentclass[twocolumn,showpacs,preprintnumbers,amsmath,amssymb]{revtex4}
\usepackage{graphicx}% Include figure files
\usepackage{dcolumn}% Align table columns on decimal point
\usepackage{bm}% bold math

\begin{document}

%\preprint{Phys. Rev. D \textbf{77}, 103012 (2008)}

\title{Resurvey of order and chaos in spinning compact binaries} % Force line breaks with \\

\author{Xin Wu$^{1}$}
\email{xwu@ncu.edu.cn}
\author{Yi Xie$^2$}
\affiliation{1. Department of Physics, Nanchang University,
Nanchang 330031, China \\ 2. Department of Astronomy, Nanjing
University, Nanjing 210093, China}

%%\date{\today} % It is always \today, today,
              %  but any date may be explicitly specified

\begin{abstract}
This paper is mainly devoted to applying the \emph{invariant} fast
Lyapunov indicator to clarify some doubt regarding the apparently
conflicting results of chaos in spinning compact binaries at the
second order post-Newtonian approximation of general relativity
from previous literatures. It is shown with a number of examples
that no single physical parameter or initial condition can be
described as responsible for causing chaos, but a complicated
combination of \emph{all} parameters and initial conditions. In
other words, a universal rule for the dependence of chaos on each
parameter or initial condition cannot be found in general. For
details, chaos does not depend only on the mass ratio, and the
maximal spins do not necessarily bring the strongest effect of
chaos. Additionally, chaos does not always become drastic when the
initial spin vectors are nearly perpendicular to the orbital
plane, and the alignment of spins cannot trigger chaos by itself
alone.
\end{abstract}

\pacs{04.25.Nx, 05.45.Jn, 95.10.Fh, 95.30.Sf}% PACS, the Physics and Astronomy
                             % Classification Scheme.
% Nonlinear dynamics and chaos, Chaotic dynamics,  Galactic nuclei (including black holes),
% circumnuclear matter, and bulges
%\keywords{Suggested keywords}%Use showkeys class option if keyword
                              %display desired
\maketitle

\section{Introduction}

Inspiralling compact binaries, made of neutron stars and/or black
holes, are the most promising sources for gravitational-waves
detectors such as LIGO and VIRGO. The successful analysis of the
experimental data requires that signals should be drawn out of a lot
of instrumental noise by matching observational data with a bank of
theoretical templates. This is a \emph{match-filtering} technique.
However, the onset of chaos would make the implementation of this
technique impractical. In this sense, the dynamical behavior of the
system becomes a fascinating and interesting topic. There have been
a number of articles in this field [1-13].

A point to deserve a little attention is that there are distinct
results on chaos and order in spinning compact binaries in
existing references [1,2,5,6,7,9,10,11]. These results seem to be
cloudy and confusing, even apparently conflictive. They are
originated from different approximations to the relativistic
two-body problem, methods for diagnosing chaos, dynamical
parameters and initial conditions. To clarify these doubt, we
shall introduce more information for each case.

(i) Different equations of motion. A detailed discussion to the
problem was given by Levin [12]. A conservative binary system
composed of a nonspinning black hole and a spinning companion have
three approximations: (1) the full relativistic system with the
extreme-mass-ratio limit of a spinning particle orbiting a
Schwarzschild black hole [1], (2) the post-Newtonian (PN)
Lagrangian formulation with one body spinning [2,8], and (3) the
PN Hamiltonian formulation with one body spinning [10,11].
Clearly, case (2) with case (3) just approaches to case (1) when
the  mass ratio becomes extreme. In spite of that, there are
differently dynamical behaviors among the three approximations.
The leading two approximations exhibit chaos [1,2], and the third
does not at 2PN order [10,11]. As an illustration, chaos in case
(1) happens only when the test particle around the Schwarzschild
black hole has an unphysically large spin. For comparable mass
binaries, either the Lagrangian formulation or the Hamiltonian
formulation is often considered. It is worth emphasizing that the
two approaches are only approximately equal, but not exactly. A
typical difference between them lies in that the constants of
motion are approximately accurate to some specific PN order for
the former, while they are exactly conserved for the latter.
Although there are such slight differences in the constants of
motion between the two formulations, there is completely different
dynamics in some sense. For instance, both the binary consisting
of comparable-mass compact objects with one physically spinning
body and the binary consisting of equal-mass compact objects with
two arbitrary spins in the 2PN Lagrangian formulation do admit
chaos [2,6-8], but their counterparts in the 2PN Hamiltonian
formulation do not because they are actually integrable in the two
simplified cases [10,11]. In all other situations, both the 2PN
Lagrangian and the 2PN Hamiltonian favor the existence of chaos.

(ii) Indicators of chaos. There have been many methods to
distinguish between ordered and chaotic motion. A Poincar\'{e}
surface of section is easy to quantify chaos when the number of
the dimension of phase space minus the number of all constants of
motion is not more than three. The Poincar\'{e} surface is
obtained by plotting a point in a certain two-dimensional (2D)
plane of phase space each time when the orbit crosses a surface on
which the other coordinates are fixed. A smooth curve, composed of
the collection of points, represents the regular motion. On the
other hand, the motion is chaotic if the points fill an area in
this plane. However, this method of identifying chaos is less than
ideal to describe the dynamics of a higher dimensional phase
space. In this case, the largest Lyapunov exponent, as a measure
of the average exponential deviation of two nearby orbits, is
often used. In classical physics, there are two different
techniques concerning its calculations. A rigorous method is
called the variational method with a tangent vector, a solution of
the variational equations [14]. It is necessary to rescale the
size of the tangent vector from time to time so that overflows can
be avoided. As an emphasis, it is cumbersome to derive the
variational equations in general. In view of this factor, an
alternative procedure to the variational method is to use the
distance $d(t)$ at time $t$ in the phase space between two nearby
trajectories as an approximation to the norm of the tangent
vector. This approach is named as the two-particle method [14].
The method is valid if initial separation d(0) is sufficiently
small and the renormalization is sufficiently frequent. By
plotting $\ln\chi(t)$ vs $\ln t$ with
$\chi(t)=(1/t)\ln[d(t)/d(0)]$, one can see that a negative
constant slope signals the regularity of the orbit. If the slope
tends gradually to zero, identically, $\ln\chi(t)$ arrives nearly
at a certain stabilizing value, the bounded system turns out to be
chaotic. This is practically attributed to a \emph{limit} method
for getting reliable Lyapunov exponents. On the other hand, there
is a slightly different treatment by plotting $\ln[d(t)/d(0)]$ vs
$t$ instead of $\ln\chi(t)$ vs $\ln t$. At once, $\chi$, as the
slope of the fit line $\ln[d(t)/d(0)]=\chi t$, is the largest
Lyapunov exponent. Here, it is necessary to perform a
least-squares fit on the simulation data. This is a \emph{fit}
method. There should have been no difference in the computation of
Lyapunov exponents between the fit method and the limit method,
but it is easier and faster to get a fit slope than a stabilizing
limit value [15]. It should be noted that a sufficiently long
integration time is needed to get reliable Lyapunov exponents,
especially for the limit method. Under this circumstance, a
quicker and more sensitive indicator, the fast Lyapunov indicator
of Froeschl\'{e} and Lega (FLI) [16], is recommended. This
indicator means the natural logarithm of the length of a deviation
vector, which stretches exponentially with time for a chaotic
orbit but linearly for an ordered orbit. Thus, it allows one to
distinguish between the ordered and chaotic cases. In addition,
the frequency analysis method of Laskar [17] is much faster to
detect chaos from regular than the method of the Lyapunov
exponent. Of course, there are other equally fast, or faster
methods to find chaos, for example, the method of the dynamical
spectra of Voglis et al. [18], and the method of the smaller
alignment index of Skokos [19]. For more information, see a
comparison of the various methods in pages 277-280 of the book
entitled ``Order and Chaos in Dynamical Astronomy" by Contopoulos
[20]. These methods, except the method of the Poincar\'{e}
surface, are independent of the dimension of phase space.

General relativity has a time-redefinition ambiguity that allows any
chaos apparently to be defined away by virtue of a spacetime
coordinate transformation. There is a long history about the
reliability of Lyapunov exponents in a curved space. A typical
example is that Lyapunov exponents in the mixmaster cosmology depend
on the choice of time variable [21-24]. Thus it is important enough
to find a gauge invariant measure of chaos. Several independent
groups have managed to work out this problem. Imponente \& Montain
[25] projected a geodesic deviation vector in the Jacobi metric on
an orthogonal tetradic basis and obtained positive Lyapunov
exponents of the mixmaster cosmology. So did Motter [26], who
addressed directly the issue of the invariance of Lyapunov
exponents. These invariant definitions of Lyapunov exponents are
mainly focused on the time evolution of the gravitational field
itself. On the other hand, for the geodesic or nongeodesic motion of
particles in a given relativistic gravitational field, Wu \& Huang
[27] gave an invariant definition of Lyapunov exponents by  refining
the classical two-particle method. In their method, space projection
operations are adopted, and coordinate time is chosen as an
independent variable. This technique works well in the study of the
chaotic dynamics in a superposed Weyl spacetime [28]. Wu et al. [29]
introduced another invariant two-particle approach of Lyapunov
exponents without projection operations and with proper time as the
independent variable for a geodesic flow. If proper time and
coordinate time do not satisfy an approximately logarithmic
relation, the two kinds of invariant two-particle methods should be
equivalent [29]. They also constructed the invariant FLI of two
nearby trajectories. As to a spinning compact binary system with the
PN equations of motion in coordinate time, it belongs to the
nongeodesic case. By the 2PN bodies' metrics in the center-of mass
(CM) frame [30-33], one can easily note that the former invariant
two-particle approach to Lyapunov exponents and its corresponding
invariant FLI become possible applications in the dynamics of this
system. Besides these invariant indicators, the fractal basin
boundary method is regarded as to a coordinate invariant approach
that gives a conclusion no ambiguity of spacetime coordinates.
Although the system in the  2PN Lagrangian formulation is
conservative, the coalescence of the black holes may occur. It
should be emphasized that the coalescence is not a consequence of
energy loss but just that these chaotic orbits happen to veer too
close at some stage and merge. In this sense, the method of fractal
basin boundaries is still a suitable tool. Usually, a basin is a 2D
space $(\theta_{1}, \theta_{2})$, where $\theta_{1}$ and
$\theta_{2}$ are two initial spin angles. In this basin $\theta_{1}$
and $\theta_{2}$ are varied, and the other initial variables are
fixed. As a result, one can determine whether the resulting orbits
coalesce, escape to infinity, or remain stable and bounded. By
color-coding each behavior and drawing points in the $(\theta_{1},
\theta_{2})$ plane, one does observe the onset of chaos if the basin
boundaries are fractal [2-4,6-8]. However, the fractal method has
some limitations; for example, it makes no distinction between
chaotic and ordered bound stable orbits in a non-fractal region of
the basin, and gives no indication of the chaotic timescale either.

Now let us recall the existing references [2-8] on this chaos
problem in the conservative 2PN Lagrangian formulation of spinning
compact binaries by means of the related qualitative techniques
above. Because of different methods adopted, there were initially
some debates about the presence or absence of chaos in the system.
With the help of the fractal basin boundary method, an earlier
paper of Levin [2] depicted that spinning compact comparable mass
binaries are shown to be chaotic in the PN expansion of the
two-body system for some range of parameters. Furthermore, the
authors of [5] employed the classical limit method to calculate
Lyapunov exponents along the fractal of Ref. [2] and presented
contradictory results. As they claimed, ``Varying the binary mass
ratio, spin magnitudes, misalignment angles, eccentricity, and
initial separation over wide ranges, we consistently find the same
regular, nonchaotic behavior for all trajectories". In brief, a
main result of [5] is that chaos in compact binary systems should
be ruled out. Cornish \& Levin [6,7] refuted the claims by showing
that the 2PN equations of motion do admit orbits near the boundary
with positive Lyapunov exponents. They explained the disagreement
between their results and those in Ref. [5], and pointed out that
``The reason for the discrepancy seems to be that the authors of
[5] define the maximum exponent as `the Cartesian distance between
the dimensionless 12-component coordinate vectors $\cdots$ of two
nearby trajectories'. $\cdots$, the result does depend on the
rescaling and can give false answers". Recently, Wu \& Xie [15]
did not think that the rescaling is an exact source of the
incorrect result that there were no positive Lyapunov exponents
and the corresponding false conclusion that chaos was ruled out in
Ref. [5]. In fact, the method for computing Lyapunov exponents in
Refs. [6,7] is slightly different from that in Ref. [5]. Refs.
[6,7] deal with the fit method in the Newtonian frame. As stated
above, the fit method is greatly superior to the limit method in
the speed of identifying chaos. Wu \& Xie [15] described that an
integration time problem is regarded as the source of the
erroneous null results in Ref. [5]. For details, the limit method
would get unreliable Lyapunov exponents in a short integration
time. Especially for coalescing binaries, the limit method is no
longer a good tool to identify chaos. On the contrary, the fit
method can find chaos in the 2PN Lagrangian approximation.
Similarly, Hartl \& Buonanno [9] applied the same method to
confirm the existence of chaos in the 2PN Hamiltonian formulation
through positive Lyapunov exponents. For an illustration, the
Lyapunov exponents calculated in [15] are taken from the
\emph{invariant} two-particle method [27] and should not have any
possible ambiguity from coordinates. Meanwhile, chaos in the
Lagrangian approximation was again confirmed using the
\emph{invariant} FLI of two nearby orbits in a curved spacetime
[15]. Additionally, chaos in this formulation was found with the
frequency map analysis [13]. In a word, it can be concluded from
[15] that chaos seems not to be ruled out in real binaries.

(iii) The choice of dynamical parameters and initial conditions.
As stated above, spinning compact binaries have 12 degrees of
freedom containing a 3D position, a 3D velocity and two 3D spins.
In addition, several parameters to affect the dynamics are: mass
ratio, magnitudes of spins, spin alignments with respect to the
orbital plane, eccentricity of orbit, and radius of orbit. Some
effects of the parameters on the dynamics were discussed in Refs.
[2,8,9,13]. Main results are listed here. (1) Levin [8] claimed
that the mass ratio primarily affects the cone of precession. A
smaller mass ratio means a wider precessional angle. Meantime the
author pointed out that it is unclear whether the mass ratio
impacts the regularity of motion. On the other hand, Hartl \&
Buonanno [9] surveyed some mass configurations, such as
$(20+25)M_{\odot}$, $(10+10)M_{\odot}$, $(20+10)M_{\odot}$ and
$(15+5)M_{\odot}$. They described that chaotic orbits occur only
for the $(10+10)M_{\odot}$ and $(20+10)M_{\odot}$ cases. (2) The
transition to chaos occurs as the spin magnitudes and
misalignments are increased. In particular, the binaries become
dramatically chaotic when the spins are perpendicular to the
orbital angular momentum [8]. This can also be seen in Ref. [13].
But there is an entirely different opinion. As shown in Ref. [9],
chaotic orbits are mainly concentrated on initial spin vectors
nearly anti-aligned with the orbital angular momentum for the
$(10+10)M_{\odot}$ configuration, while they are located at other
initial spin directions for the $(20+10)M_{\odot}$ configuration.
(3) Levin [8] found that large eccentricity  does not cause chaos
alone. On the contrary, Hartl \& Buonanno [9] did think that chaos
appears in highly eccentric orbits. In sum, it can be concluded
that there are completely different and apparently conflicting
descriptions of chaotic regions and parameters to the binaries
between in Ref. [8] and in Ref. [9].

As mentioned above, the debates in (i) and (ii) have been given
satisfactory answers by several authors, but those in (iii) have not
yet. Thus, an important motivation of the present paper is to
clarify these doubt regarding chaotic regions and parameters to
black hole pairs in Refs. [8] and [9]. In our opinion, the reason
for the apparently conflicting results is that each of physical
parameters or initial conditions is taken solely to be responsible
for causing chaos in the two references. We do believe that all
these results should not conflict as long as a complicated
combination of all parameters and initial conditions can be adopted
as a criterion for chaos. For a representative example to argue
these points of view, we consider only the 2PN Lagrangian
formulation of spinning compact binaries. We continue to trace chaos
and order in this model with the invariant FLI along the previous
work [15]. On one hand, the superiority of this indicator in the
application is described sufficiently so that we can take the
opportunity to examine the method of fractal basin boundaries
adopted in a series of articles [2,6-8,12] about chaos in this
formulation. On the other hand, we shall focus on the transition to
chaos with one or two of parameters varied. As an emphasis, we are
interested in investigating the regularity or chaoticity of stable
orbits within the integration time considered rather than that of
unstable orbits at the basin boundaries. Above all, some details
neglected in the existing references will be concerned. For example,
we shall study whether chaos depends on the mass ratio, and also
wonder whether the maximal spin magnitudes do always increase the
strength of the chaotic behavior in any term.

The paper is structured in the following manner. In Sec. II, we
exhibit the related invariant chaos indicators in spinning compact
binaries. Then we use the invariant FLI to explore the effects of
various parameters on the dynamical transition to chaos in Sec.
III. Finally, the summary follows in Sec. IV. Throughout the work
we use geometric units $c=G=1$, and take the signature of a metric
as $(-,+,+,+)$. Greek subscripts run from 0 to 3, and Latin
indexes range from 1 to 3.

\section{Invariant indicators for identifying chaos in black hole pairs}
Invariant chaos indicators means that they should be independent
of the choice of spacetime coordinates for a given relativistic
dynamical problem. This is a basic requirement of full general
relativity. In order to construct the invariant chaos indicators
in black hole binaries, we need the bodies' motions and metrics in
the CM frame. For this purpose, we list some basic characteristics
of spinning compact binaries, for example, the equations of the
relative motion, the relation between the relative motion and the
bodies' motions, and the bodies' metrics. Then, we introduce both
the invariant Lyapunov exponent and the invariant FLI of two
nearby trajectories.

\subsection{Equations of the relative motion}
For a relativistic system of two pointlike particles with masses
$m_1$ and $m_2$ ($m_1\geq m_2$), and the total mass $M=m_1+ m_2$,
the relative position $\mbox{\boldmath$x$}$ and velocity
$\mbox{\boldmath$v$}$ from body 2 to body 1 evolve according to
the Lagrangian formulation at 2PN order in harmonic coordinates:
\begin{equation}
\ddot{\mbox{\boldmath$x$}}=\mbox{\boldmath$a$}^{(0)}_{N}
+\mbox{\boldmath$a$}^{(1)}_{PN}
+\mbox{\boldmath$a$}^{(1.5)}_{SO}+\mbox{\boldmath$a$}^{(2)}_{PN}
+\mbox{\boldmath$a$}^{(2)}_{SS}.
\end{equation}
The explicit forms of $\mbox{\boldmath$a$}$ can be found in Ref.
[34]. The superscripts stand for the order of the PN expansion,
and the subscripts represent the type of the contributions to the
relative acceleration, which are from the Newtonian (N) and
post-Newtonian (PN), and the spin-orbit (SO) and spin-spin (SS)
coupling. In addition, the two spins satisfy
\begin{equation}
\dot{\mbox{\boldmath$S$}}_{1}=\mbox{\boldmath$\Omega$}_{1} \times
\mbox{\boldmath$S$}_{1}, ~~~~~~
\dot{\mbox{\boldmath$S$}}_{2}=\mbox{\boldmath$\Omega$}_{2} \times
\mbox{\boldmath$S$}_{2},
\end{equation}
with
\begin{eqnarray}
\mbox{\boldmath$\Omega$}_{1} = \frac{1}{r^{3}}
[(2+\frac{3\beta}{2}) \mbox{\boldmath$L$}_{N}
-\mbox{\boldmath$S$}_{2}
+3(\mbox{\boldmath$n$}\cdot\mbox{\boldmath$S$}_{2})\mbox{\boldmath$n$}],
& &
\nonumber \\
 \mbox{\boldmath$\Omega$}_{2} = \frac{1}{r^{3}}
[(2+\frac{3}{2\beta}) \mbox{\boldmath$L$}_{N}
-\mbox{\boldmath$S$}_{1}
+3(\mbox{\boldmath$n$}\cdot\mbox{\boldmath$S$}_{1})\mbox{\boldmath$n$}].
& &
\end{eqnarray}
Here mass ratio $\beta=m_2/m_1$, the Newtonian orbital angular
momentum $\mbox{\boldmath$L$}_{N}=\mu(\mbox{\boldmath$x$}
\times\mbox{\boldmath$v$})$ with the reduced mass $\mu=m_1m_2/M$,
radius $r=|\mbox{\boldmath$x$}|$ and unit radial vector
$\mbox{\boldmath$n$} =\mbox{\boldmath$x$}/r$. Thus Eqs. (1) and (2)
do completely determine the evolution of the relative one-body
problem with 12 degrees of freedom in the phase space. Eq. (2)
implies that the individual spin magnitudes, $S_1$ and $S_2$, are
always constants of motion. For physically realistic spins, two spin
magnitudes are $S_1=\chi_1 m^{2}_{1}$ and $S_2=\chi_2 m^{2}_{2}$
with dimensionless spin parameters $0\leq\chi_1, \chi_2 \leq 1$.
There are also other quantities conserved at 2PN order as follows:
the total energy $E$ and the angular momentum $\mbox{\boldmath$J$}=
\mbox{\boldmath$L$}+ \mbox{\boldmath$S$}_{1}
+\mbox{\boldmath$S$}_{2}$, where $\mbox{\boldmath$L$}$ is the
orbital angular momentum.

It is worth noting that the relative coordinate
$\mbox{\boldmath$x$}$ is no other than a separation between the
body coordinates $\mbox{\boldmath$y$}_1$ and
$\mbox{\boldmath$y$}_2$ in the CM frame, namely,
$\mbox{\boldmath$x$}= \mbox{\boldmath$y$}_1-
\mbox{\boldmath$y$}_2$. Meantime, $\mbox{\boldmath$v$}=
\mbox{\boldmath$v$}_1- \mbox{\boldmath$v$}_2$, where
$\mbox{\boldmath$v$}_1$ and $\mbox{\boldmath$v$}_2$ denote the
body velocities. Inversely, the relative motion can determine the
motion of each body. In other words, both $\mbox{\boldmath$y$}_1$
and $\mbox{\boldmath$y$}_2$ can be given by $(\mbox{\boldmath$x$},
\mbox{\boldmath$v$})$. For details, see the next subsection.

\subsection{Center-of-mass coordinates}

Besides the related parameters above, we define mass parameters
$\eta=m_1m_2/M^2$ and $\delta m=(m_1-m_2)/M$, and dimensionless
spin parameters [31]
\begin{eqnarray}
\mbox{\boldmath$\chi$}_{+} &=& (\mbox{\boldmath$S$}_{1}/m^{2}_{1}+
\mbox{\boldmath$S$}_{2}/m^{2}_{2})/2, \\
\mbox{\boldmath$\chi$}_{-} &=& (\mbox{\boldmath$S$}_{1}/m^{2}_{1}-
\mbox{\boldmath$S$}_{2}/m^{2}_{2})/2.
\end{eqnarray}
The 2PN-accurate relationship between the individual CM
coordinates $\mbox{\boldmath$y$}_{1}$ and
$\mbox{\boldmath$y$}_{2}$, and the relative variables
$(\mbox{\boldmath$x$},\mbox{\boldmath$v$})$  is written as
\begin{eqnarray}
\mbox{\boldmath$y$}_{1} &=& (m_2/M +\eta \delta m \mathcal{P})
\mbox{\boldmath$x$} +\eta \delta m
\mathcal{Q}\mbox{\boldmath$v$} \nonumber \\
& & -M\eta \mbox{\boldmath$v$}\times (\mbox{\boldmath$\chi$}_{+}
+\delta m
\mbox{\boldmath$\chi$}_{-}), \\
\mbox{\boldmath$y$}_{2} &=& (-m_1/M +\eta \delta m \mathcal{P})
\mbox{\boldmath$x$} +\eta \delta m
\mathcal{Q}\mbox{\boldmath$v$} \nonumber \\
& & -M\eta \mbox{\boldmath$v$}\times (\mbox{\boldmath$\chi$}_{+}
+\delta m \mbox{\boldmath$\chi$}_{-}).
\end{eqnarray}
The last terms in the above equations are of the 1.5PN terms given
by Ref. [31], while $\mathcal{P}$ and $\mathcal{Q}$ presented in
Ref. [32] contain the 1PN and 2PN terms of the type
\begin{eqnarray}
\mathcal{P} &=& (\frac{v^2}{2}-\frac{M}{2r})
+[\frac{3}{8}v^4-\frac{3}{2}\eta v^4 \nonumber \\
& & +\frac{M}{r}(-\frac{\dot{r}^2}{8} +\frac{3}{4}\eta\dot{r}^2
+\frac{19}{8}v^2 +\frac{3}{2}\eta
v^2) \nonumber \\
& & + \frac{M^2}{r^2}(\frac{7}{4}-\frac{\eta}{2})], \\
\mathcal{Q} &=& -7M\dot{r}/4,
\end{eqnarray}
where the relative velocity magnitude $v=|\mbox{\boldmath$v$}|$
and the radial velocity $\dot{r}=\mbox{\boldmath$n$}\cdot
\mbox{\boldmath$v$}$. As to $\mbox{\boldmath$v$}_1$ and
$\mbox{\boldmath$v$}_2$, they are  from derivatives of
$\mbox{\boldmath$y$}_{1}$ and $\mbox{\boldmath$y$}_{2}$ with
respect to coordinate time $t$, respectively, where all the terms
higher than 2PN order are dropped. It should be pointed out that
each body has its metric that governs its evolution.

\subsection{Metric of body 1 in the CM frame}

In the light of the body coordinates $\mbox{\boldmath$y$}_1$ and
$\mbox{\boldmath$y$}_2$ and the body velocities
$\mbox{\boldmath$v$}_1$ and $\mbox{\boldmath$v$}_2$, Faye et al.
[33] provided the 3PN harmonic-coordinate metric coefficients
computed at body 1 in the CM frame. As mentioned above, here we
remain them to the 2PN order. They take the following forms
\begin{eqnarray}
g_{00} &=& -1+2V-2V^2 \nonumber \\
& & +8(\hat{X} +V_iV_i+ \frac{V^3}{6}), \\
g_{0i} &=& -4V_i-8\hat{R}_i, \\
g_{ij} &=& \delta_{ij} (1+2V+2V^2)+4\hat{W}_{ij}.
\end{eqnarray}
Each of all the potentials is split into the non-spin (NS) piece
given by Ref. [30] and the spin (S) part listed in Ref. [31], say
\begin{eqnarray}
V &=& V_{NS}+V_{S}, \cdots, \\
\hat{W}_{ij} &=& \hat{W}_{ij;NS}+\hat{W}_{ij;S}.
\end{eqnarray}
Additionally, the proper time $\tau$ of body 1 satisfies the
following equation
\begin{equation}
\frac{d\tau}{dt}=\sqrt{-(g_{00}+2g_{0i}v^{i}_{1}
+g_{ij}v^{i}_{1}v^{j}_{1})}.
\end{equation}
The superscript $i$ denotes the $i$th component of the velocity for
body 1. Then body 1 has its 4-velocity
\begin{equation}
\mbox{\boldmath$U$}=(\frac{dt}{d\tau}, v^{1}_{1}\frac{dt}{d\tau},
v^{2}_{1}\frac{dt}{d\tau}, v^{3}_{1}\frac{dt}{d\tau}).
\end{equation}

In terms of Eqs. (6) and (7), the dynamical behaviors of the
bodies' motions should be equivalent to those of the relative
motion. This requires that chaos indicators should be  invariant
for various spacetime coordinates. With the aid of the metric
$g_{\alpha\beta}$, the invariant Lyapunov exponent of two nearby
trajectories [27] can be constructed and used to study the
dynamics of orbits around body 1.

\subsection{The invariant Lyapunov exponent}
According to the theory of observation in general relativity, Wu
\& Huang [27] employed proper time $\tau$ of an ``observer" and a
proper configuration space distance $\Delta L (\tau)$ between the
observer and his ``neighbor" particles to define an
\emph{invariant} Lyapunov exponent:
\begin{equation}
\lambda = \lim_{\tau \rightarrow \infty}\chi(\tau),
\end{equation}
where
\begin{equation}
\chi(\tau) =\frac{1}{\tau} \ln\frac{\Delta L(\tau)}{\Delta L(0)},
\end{equation}
with the proper distance between the two nearby trajectories
\begin{equation}
\Delta L (\tau)= \sqrt{h_{\alpha \beta} \Delta x^{\alpha}\Delta
x^{\beta}}.
\end{equation}
In addition, let the space projection operator of the observer be
$h^{\alpha\beta}= g^{\alpha\beta}+U^{\alpha}U^{\beta}$, and the
deviation vector from the observer to the neighbor be $\Delta
x^{\beta}$.

In fact, this technique is no other than a directly modified and
refined version of the classical Lyapunov exponent with two nearby
trajectories [14]. A point to note is that the coordinate time $t$
still remains of a common time variable in the equations of motion
for the two particles, but the proper time $\tau$ is from
integration of Eq. (15). As $\tau\sim\ln t$, this method is invalid
[29]. But this case does not appear in spinning compact binaries.

On the other hand, the invariant Lyapunov exponent is not suitable
for the study of comparable mass compact binaries with spins because
of the merger, as mentioned in Ref. [15]. Thus, we shall
particularly focus on the application of the invariant FLI.

\subsection{The invariant fast Lyapunov indicator}

The so-called Fast Lyapunov Indicator (FLI) of Froeschl\'{e} \&
Lega [16] has been widely used to survey various orbital problems,
\emph{e.g.}, see Ref. [35]. However, there is generally great
difficulty in deriving the variational equations corresponded to
the tangential vector for complicated problems, especially for
relativistic gravitational systems. To avoid this, Wu et al. [29]
refined the original idea and proposed the invariant FLI with two
nearby trajectories, where a renormalization technique within a
sufficiently long time span is adopt.

Body 1, as an observer, uses the above proper distance $\Delta L$
to his neighboring orbit at his proper time $\tau$ to measure the
invariant FLI of two nearby trajectories in a curved spacetime,
defined as
\begin{equation}
 FLI(\tau) = \log_{10}\frac{\Delta L(\tau)}{\Delta L(0)},
\end{equation}
where $\Delta L(0)=10^{-9}$ is an ideal choice of the starting
proper distance [29]. By plotting $ FLI(\tau)$ vs $\log_{10}
\tau$, one can know that the exponential stretching indicates the
onset of chaos, while the linear growth turns out to be regular.
Ref. [29] gives the numerical setup of this indicator in the
following.

Utilizing a fifth-order Runge-Kutta-Fehlberg algorithm of an
adaptive coordinate time step, we numerically integrate the
equations (1), (2) and (15) together two times with two groups of
slightly different initial conditions. In other words, the
coordinate time $t$ is taken as a common integration time variable
in connect with the relative motion and the bodies' motions, and
the numerical integration is used to solve the equations of the
relative motion. But the relativistic dynamics is investigated in
the CM frame, and whether chaos or not is measured by body 1. An
important point to note is that the saturation of bounded chaotic
orbits appears when $\Delta L=1$. For the sake of its
disappearance, the rescaling is not introduced until $\Delta L$
reaches the value of 0.1. Let $k ~(k=0,1,2,\cdots)$ be the
sequential number of renormalization, then a detailed algorithm of
the FLI is
\begin{equation}
FLI_{k} = -k [1+\log_{10} \Delta\mbox{\boldmath$L$}(0)] +
\log_{10} \frac{\Delta\mbox{\boldmath$L$}(\tau)}
{\Delta\mbox{\boldmath$L$}(0)},
\end{equation}
where $\Delta\mbox{\boldmath$L$}(0) \leq
\Delta\mbox{\boldmath$L$}(\tau) \leq 0.1$.

Zhu et al. [36] attained a success in the analysis of the dynamics
of Newtonian core-shell systems with the FLI of two nearby orbits.
In addition, we had already applied the \emph{invariant} FLI to
conduct a first-step investigation into the dynamics of spinning
compact binaries in [15]. Next, we shall continue to discuss its
applications in this problem.

\section{Applications of the invariant FLI}

Following Ref. [15], we employ the invariant FLI to give a
detailed discussion on the transition of the dynamics of spinning
compact binaries from regular motion to chaos with variations of
dynamical parameters or initial conditions. In particular, we are
engaged to our scan to search for chaos as initial spin angles are
varied.

To illustrate the use of the invariant FLI, we begin by
regenerating Fig. 3 of Ref. [15] in our Fig. 1, where the FLIs of
three orbits vary with the proper time. Initial conditions and
parameters of the three orbits are as follows. $\Gamma_1$:
$(\mbox{\boldmath$x$}, \mbox{\boldmath$v$})=(5.5M,0,0,0,0.4,0)$,
$\beta=1/3$, $\chi_{\imath}=1 ~(\imath=1,2)$, and spin angles
$\theta_{1}=\pi/2$ and $\theta_{2}=\pi/6$ so that initial spin
configurations $\mbox{\boldmath$S$}_{\imath}
=(S_{\imath}\sin\theta_{\imath},0,
S_{\imath}\cos\theta_{\imath})$. $\Gamma_2$:
$(\mbox{\boldmath$x$},\mbox{\boldmath$v$})=(5.0M,0,0,0,0.399,0)$,
$\beta=1$, $\chi_{\imath}=1$, $\theta_{1}=38^{\circ}$, and
$\theta_{2}=70^{\circ}$. $\Gamma_3$ is the same as $\Gamma_2$ but
only 0.399 gives place to 0.428. In practice, the initial radius
$r$ is equal to the first component $x$ of the initial relative
coordinates for each case. In order to get their corresponding
neighboring orbits, we add a very small deviation, $\Delta
x=10^{-9}M$, to the $x$ only. As shown in Fig. 1, $\Gamma_1$ and
$\Gamma_2$ are chaotic, but $\Gamma_3$ becomes ordered. The
results are consistent with those of [7]. Obviously, chaos of
$\Gamma_1$ gets rather stronger than that of $\Gamma_2$. In
particular, the three orbits can be distinguished clearly  as
proper time arrives at $10^{5}M$.

Hereafter, each orbit does not stop computing till $\tau=10^{5}M$.
It is shown with many numerical experiments that $FLI=6$ is a
threshold between ordered and chaotic at this time. In detail, all
orbits with $FLIs>6$ show chaos, while ones with $FLIs\leq6$
signal regular. A point to note is that we shall hardly consider
coalescing orbits during this time, but shall pay attention to
\emph{stable} orbits. Of course, increasing the integration time
would reduce the number of the stable orbits. Since the invariant
FLI has explicit merits, we shall borrow it to further gain an
insight into the dynamics of spinning compact binaries.

\subsection{Varying initial radii}

Let us trace a dynamical sensitivity to the variations of some
initial variables of the compact objects by taking $\Gamma_1$,
$\Gamma_2$ and $\Gamma_3$ as basic references.

Let us create a family of orbits with the same parameters and
initial conditions as those of $\Gamma_1$ but the initial radius
$r$ altered from $5.4M$ to $6M$. In terms of different values of
FLIs, Fig. 2a describes that chaotic orbits are clustered at small
radius regions of $r$ lower than $5.58M$. When different initial
conditions and parameters $\dot{y}$, $\theta_1$, $\theta_2$ and
$\beta$ are used, one can see the same fact from Figs. 2b-2f.
There are two main reasons for leading to the result. A smaller
possible radial separation corresponds to greater contributions to
the PN terms in the equations of motion so that the nonlinear
effects become stronger. On the other hand, it gives rise to the
stronger spin coupling, as hinted in Eq. (3).

For an illustration, the presence of chaos at low radius regions
was already mentioned by Hartl \& Buonanno [9], who used another
approximation, the PN \emph{ADM-Hamiltonian} formulation for the
same problem.

\subsection{Varying initial eccentricities}

Now, we observe the dynamical evolution with second initial
component $\dot{y}$ of the relative velocity vector
$\mbox{\boldmath$v$}$ running from the interval [0.39, 0.633]
according to Fig. 3a, where a number of orbits  are consistent
with $\Gamma_2$ in the parameters and the initial conditions
except the $\dot{y}$. Without question, $\Gamma_2$ and $\Gamma_3$
are still two of all the orbits tested. Similar to Fig. 2, Fig. 3a
displays that the onset of chaos is also at lower velocities less
than about 0.42. In practice, the variation of $\dot{y}$ at the
starting time corresponds to that of initial eccentricity when all
the parameters and the other initial conditions are fixed.
$\dot{y}=0.44721$ means the initial eccentricity $e=0$, which just
corresponds to a quasicircular orbit of Newtonian two-body
problems. Of course, the chaoticity of quasicircular orbits in
relativistic  two-body problems is possible for particular
parameters and initial conditions (see Ref. [9]). In addition, $e$
becomes small with the growth of $\dot{y}$ when $\dot{y}<0.44721$
or large with the increase of $\dot{y}$ when $\dot{y}>0.44721$. In
particular, we have $e=1$ for $\dot{y}=0.63245$. In addition,
$e=0.2$ for $\Gamma_2$, while $e=0.08408$ for $\Gamma_3$. For
details, see Fig. 3b, which plots $FLI$ vs initial eccentricity
$e$ rather than $FLI$ vs $\dot{y}$ in Fig. 3a. Clearly, chaotic
orbits are mainly concentrated on $e\approx0.2$. However, Fig. 3c
shows that chaos does occur near $e=0$ when we change the initial
radius and spin angles, i.e. $r=5.3M$, $\theta_1=\theta_2=\pi/2$.
On the other hand, it can still be seen from Fig. 3d  that chaotic
orbits are approximately located in $e=0.2$ when $r=5.5M$,
$\beta=1/3$, $\theta_1=0.91$ and $\theta_2=2.45$. Of course, the
relation for chaos dependence of initial eccentricity should vary
if the parameters and the other initial conditions are different.

Meanwhile, Fig. 3 seems to tell us that any highly initial
eccentricity does not bring chaos. Here we provide some details of
the choice of the initial velocity $\dot{y}$ for a given initial
eccentricity. In our various cases tested, we get
$\dot{y}=\sqrt{(1+e)/r}$ or $\dot{y}=\sqrt{(1-e)/r}$ from the
equation of $e=|r\dot{y}^{2}-1|$. In general, black hole binaries do
not coalesce fast for highly initial eccentricity with larger
starting velocity of $\dot{y}=\sqrt{(1+e)/r}$. In fact, all ordered
orbits with high eccentricities in Fig. 3 are just corresponded to
this case. On the contrary,  for highly initial eccentricity with
smaller starting velocity of $\dot{y}=\sqrt{(1-e)/r}$, the merger of
black hole binaries appears so quicker that we have no way to detect
chaos from order. In this sense, we cannot say that highly initial
eccentricity with smaller starting velocity does not cause chaos. It
has been reported that highly eccentric chaotic orbits with some
particular parameters and initial conditions do exist in the
corresponding \emph{ADM-Hamiltonian} formulation [9].

The above facts show that eccentricity alone is not responsible for
causing chaos. The result is the same as that of [8].

\subsection{Varying binary mass ratio}

Levin [8] investigated the binary mass ratio $\beta$ how to affect
the bulk shape of the precession and gravitational wave
modulation. As a result, the smaller the mass ratio is, the more
prominent the effect of the precession becomes. Still, it has been
unclear how much the mass ratio impacts the appearance of chaos.
This is what we want to explore.

Hereafter we specify all orbits with initial conditions and
parameters: $x=r$, $y=z=\dot{x}=\dot{z}=0$, $\chi_1=\chi_2=1$
(except Fig. 5), and with the others marked in each panel. As
shown in Fig. 4a with $r=6M$, $\dot{y}=0.399$, $\theta_1=0$ and
$\theta_2=2.23~ radians$, the dynamics is typically ordered for
huge numbers of values of $\beta$. A little attention to deserve
is that a transition to chaos happens only when $\beta\approx1$.
However, the case is entirely different when we adopt $r=5.5M$,
$\dot{y}=0.4$, $\theta_1=\pi/2$ and $\theta_2=\pi/6$ in Fig. 4b,
which displays that chaos exists for most values of $\beta$. It is
emphasized that chaos occurs neither for the maximum mass ratio
nor for the minimum mass ratio. On the other hand, there are
differently dynamical transitions with variations of $\beta$ if
$r$, $\dot{y}$, $\theta_1$ and $\theta_2$ are changed into other
fixed values like those given by Fig. 4c and 4d.

Therefore, it can be concluded from Fig. 4 that the mass ratio,
$\beta$, can not be used \emph{only} as a criterion for causing
chaos or order. The larger or the smaller the mass ratio is, the
stronger chaos does not necessarily get. There are various cases
about the transitivity to chaos with the binary mass ratio varying
for different combinations of fixed initial conditions and other
parameters. A tentative interpretation to this thing is because
the system studied has such many degrees of freedom and parameters
that the dynamical features depend on not only $\beta$ but also
the others. Once $\beta$ is varied (but the others are fixed and
permitted to be chosen as variously possible values), it is not
surprise to see the differently dynamical behaviors above.

\subsection{Varying spin magnitudes}

Levin [8] as well as Hartl \& Buonanno [9] studied the effect of
spinning up the lighter companion with some orbits. They pointed out
that the larger the magnitude of the second spin becomes, the more
irregular the motion is. Now let us use the invariant FLI to check
this fact.

We take $\chi_1=\chi_2=\chi$, and then let $\chi$ range from 0.1
to 1. As expected, there is an abrupt transition to chaos in Fig.
5a when $\chi$ exceeds 0.86. It is worth emphasizing that chaos is
incurred dramatically as $\chi$ increases. Only when
$\theta_2=\pi/6$ in Fig. 5a gives place to $\theta_2=\pi/2$, are
chaotic orbits in Fig. 5b mainly focused on the range of $\chi$
near 0.45. Still, there remains a rather small chaotic belt close
to the maximal spins. If we set $r=5M$, $\dot{y}=0.428$,
$\theta_1=38^{\circ}$ and $\theta_2=70^{\circ}$, chaos occurs when
 $\chi$ is nearly located in the middle of the interval. More details can be
seen in Fig. 5c. On the other hand, chaos is completely absent for
any spin magnitudes when we employ $r=5.2M$ instead of $r=5M$, as
shown in Fig. 5d. With different parameters and initial conditions
adopted, the dependence of chaos on $\chi$ alters at once (see Fig.
5e and 5f).

In our opinion, the maximal spins are not necessary to bring the
strongest chaos. As stated in the above subsection, the spin
magnitudes, as one of various factors to affect chaos, are not very
sufficient to determine what dynamical feature the system is. Note
that the result in the present paper should not be in conflict with
those in Refs. [2,8], where only some particular conditions and
parameters are considered.

\subsection{Varying initial spin directions}

In this subsection, we concentrate on demonstrating some dynamical
sensitivity to the initial spin alignments. To do this, we fix the
first initial spin angle $\theta_1$ of $\mbox{\boldmath$S$}_1$,
and vary the second initial spin angle $\theta_2$ of
$\mbox{\boldmath$S$}_2$ in $[0, \pi]$. Let us choose $\beta=1/3$,
$r=6M$, $\dot{y}=0.395$ and $\theta_1=\pi/2$. It is shown with
Fig. 6a that chaos is mainly trapped in the values  of
$\theta_2\approx\pi/2$. This seems to confirm the result of Refs.
[2,8]. Of course, there is also a small chaotic region around
$\theta_2\approx\pi$. If we change only the value of $\theta_1$
and obtain $\theta_1=2$, we find in Fig. 6b that there is a
differently dynamical sensitivity to the initial spin angle
$\theta_2$. In particular, the chaotic orbits are mainly clustered
at values of $\theta_2$ around 2.25 rather than $\pi/2$ when we
take $\beta=1$, $r=6M$, $\dot{y}=0.399$ and $\theta_1=\pi/2$ in
Fig. 6c. The case is also different in Fig. 6d.

Clearly, the invariant FLI has been an invaluable and a
computationally quicker tool to survey phase space for chaos by
scanning huge numbers of orbits. Naturally, it is easy and suitable
to study the dynamical structure of the $(\theta_1, \theta_2)$
plane. Let the two initial spin angles run from $[0, \pi]$ within a
span of $\Delta\theta=0.01~ radian$, respectively. At once, we have
Figs. 7a-7d that give all the starting points in the plane according
to distinct values of FLIs so that ordered and chaotic regions can
be distinguished. In detail, the initial conditions in the
$(\theta_1, \theta_2)$ plane are color coded black if $FLIs>6$ and
gray when $FLIs\leq6$. The black indicates chaos, but the gray means
regular. By comparing between Figs. 7a and 7b or between Figs. 7c
and 7d, we find again that the dynamical structures differ greatly
for the different mass ratios. It should be pointed out that the
structures in Fig. 7b/7d are symmetric since the pairs are of the
same mass. As mentioned above, smaller initial radii lead still to
the onset of stronger chaos. In addition, it can easily be seen that
there is no chaos at all when the spins in each panel are nearly
aligned with the orbital angular momentum, i.e. at the values near
$\theta_1=\theta_2=0$. On the other hand, Eqs.  (2) and (3) seem to
show that the initial spins perpendicular to the orbital plane turn
out to have the strong effects of the spin couplings. This seems to
imply that the motion in the system becomes more irregular in this
case. In fact, there is a stronger chaotic belt around
$\theta_1=\theta_2=\pi/2$ of Fig. 7a. The result coincides basically
with that of [8]. However, an important point to note is that the
conservative system preserves only the spin magnitudes rather than
the spin directions. Therefore, it should be reasonable that there
are other chaotic regions far away values of
$\theta_1=\theta_2=\pi/2$ in Fig. 7a when different combinations of
other parameters and initial conditions are employed. Especially, it
is no surprise that chaos disappears in a neighboring region of the
point $(\pi/2, \pi/2)$ on the $(\theta_1, \theta_2)$ plane in Figs.
7c and 7d. As mentioned in the Introduction, Hartl \& Buonanno [9]
observed other cases from the PN Hamiltonian formulation, too.

In  fact, each of Figs. 7a, 7b and 7d shows fractal boundaries at
the basins of between stability (color-coded \emph{black and gray})
and merger (color-coded \emph{white}) in a slice through phase
space. As Levin [2,8,12] pointed out, the fractal basin boundaries
provide unambiguous signals of chaos. However, there is an
exceptional case in which  all the orbits of Fig. 7c are stable.
This implies that the fractal basin boundary method is no longer fit
for identifying the presence of chaos. Generally speaking, our
method for establishing chaos hints the kernel of the fractal basin
boundary method. Above all, the FLI is more universal in
application, and gives more dynamical details than the fractal basin
boundary method.

In the light of the above statements, we do not think that the
initial spins perpendicular to the orbital plane can necessarily
produce the strongest effect of chaos. There should be various
ordered and chaotic regions on the $(\theta_1, \theta_2)$ plane if
different combinations of other parameters and initial conditions
are used. It should be worth noting that the results are not
opposite to those in Refs. [2,8,9] with some particular parameters
and initial conditions.

\section{Conclusions}

For conceptual clarity, it is physically significant to apply the
invariant indicators of chaos, which are independent of the choice
of spacetime coordinates, to study the orbital dynamics of
relativistically  gravitational systems. For spinning compact
binaries, the coordinate time should be used in order to connect
the relative motion of the bodies and their internal motions. In
terms of this point and the 2PN metric of body 1 in the CM frame,
it is able to construct the invariant indicators that measure the
dynamical features of body 1, equivalently, ones of the relative
motion. In this sense, it seems to be the most preferable to adopt
the invariant Lyapunov exponent with two nearby trajectories [27].
Considering the too slow convergence of the Lyapunov exponent for
the case of comparable mass binaries, we recommend the invariant
FLI of two nearby trajectories  in a curved spacetime [29], viewed
as a very fast and valid technique to detect chaos from order.

A main contribution of the present paper is to discuss some
applications of the invariant FLI in the study of dynamical
transitions to chaos with the variations of parameters and initial
conditions for the relativistic two-body system at 2PN order. Above
all, this paper has been engaged to clarifying some doubt regarding
the apparently conflicting results of chaos in the system from
previous literatures. With this indicator we have successfully
estimated effects of varying initial radii and
velocities/eccentricities, varying binary mass ratios, varying spin
magnitudes, and varying initial spin angles on the qualitative
changes in the dynamical behaviors from nonchaotic to chaotic. For
the specific choice of parameters and initial conditions, we recover
some results of Levin [8] or Hartl \& Buonanno [9], which include:
(1) chaotic orbits are mainly clustered at initial low-radius
regions; (2) eccentricity alone is not responsible for chaos; (3)
the maximal spins increase the strength of chaos; (4) chaos becomes
drastic when the initial spin vectors are nearly perpendicular to
the orbital plane. However, when different combinations of the
dynamical parameters and initial conditions are considered, a
universal rule for the dependence of chaos on single parameter or
initial condition cannot be found in general. That is to say, chaos
does not depend only on the mass ratio. In addition, the maximal
spins do not necessarily bring the strongest chaos. On the other
hand, there are other large chaotic regions far away from the point
$(\pi/2, \pi/2)$ in the $\theta_1$-$\theta_2$ plane. Even, chaos
does disappear near the point $(\pi/2, \pi/2)$. A fundamental reason
resulting from these facts is that a spinning compact binary system
has so many degrees of freedom and parameters that only one physical
parameter or initial condition is necessary but not sufficient to
determine what dynamical behavior the system is. In short, no single
physical parameter or initial condition can be described as
responsible for causing chaos, but a complicated combination of
\emph{all} parameters and initial conditions.

It should be emphasized that the invariant FLI is a simple and firm
tool to scan the global structure of phase space of the complicated
spinning compact binary systems. Especially, the FLI is more
universal to use, and gives more dynamical details than the fractal
basin boundary method. As a result, the onset of the chaotic
behavior for these systems at the 2PN expansion has been confirmed
again.

\begin{acknowledgments}
We would like to thank the referee for the related comments and
significant suggestions. We are also grateful to Prof. Tian-Yi Huang
of Nanjing University for his helpful discussion. Dr. Xiao-Sheng Wan
provided numerical computations in part. This research is supported
by the Natural Science Foundation of China under Contract No.
10563001. It is also supported by the Science Foundation of Jiangxi
Province (0612034), the Science Foundation of Jiangxi Education
Bureau (200655), and the Program for Innovative Research Team of
Nanchang University.
\end{acknowledgments}

\begin{figure}[h]
\includegraphics[scale=0.85]{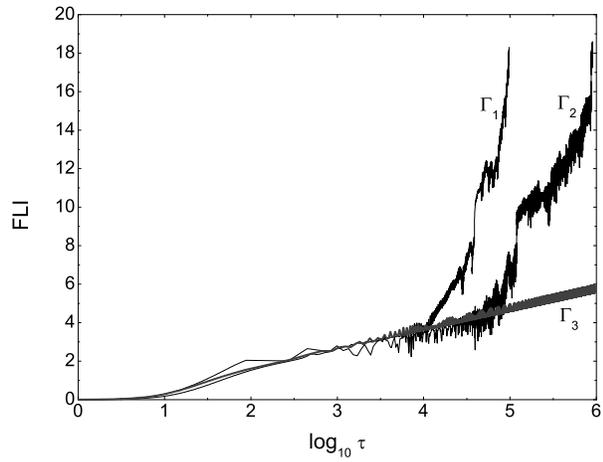}% Here is how to import EPS art
\caption{Same as Fig. 3 of Ref. [15], which describes the invariant
FLI as a function of proper time for each of three orbits. Orbits
$\Gamma_1$ and $\Gamma_2$ are chaotic, while orbit $\Gamma_3$ is
not.} \label{fig1}
\end{figure}

\begin{figure*}[h]
\includegraphics[scale=0.75]{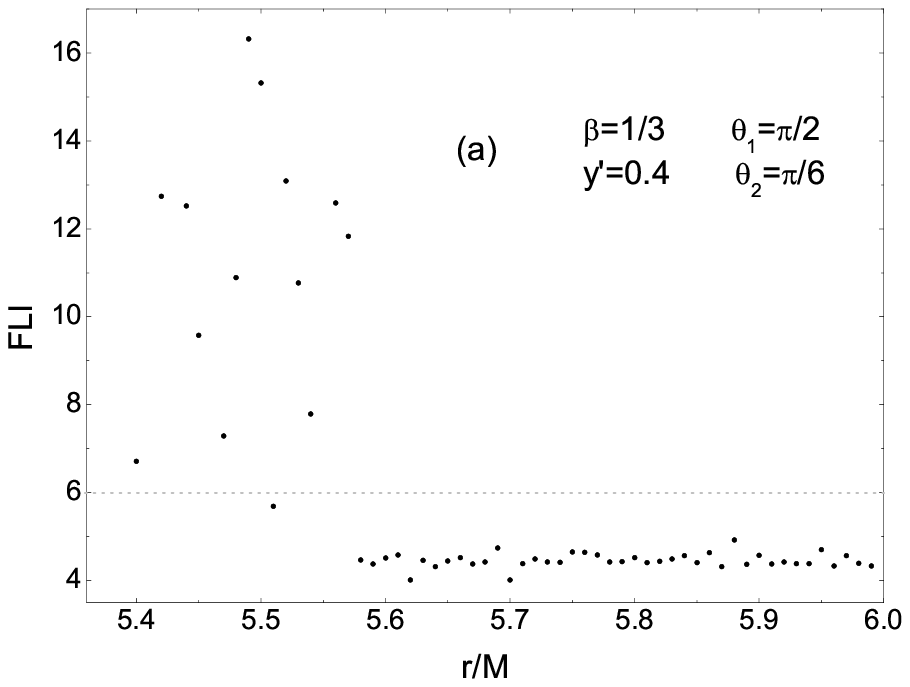}% Here is how to import EPS art
\includegraphics[scale=0.75]{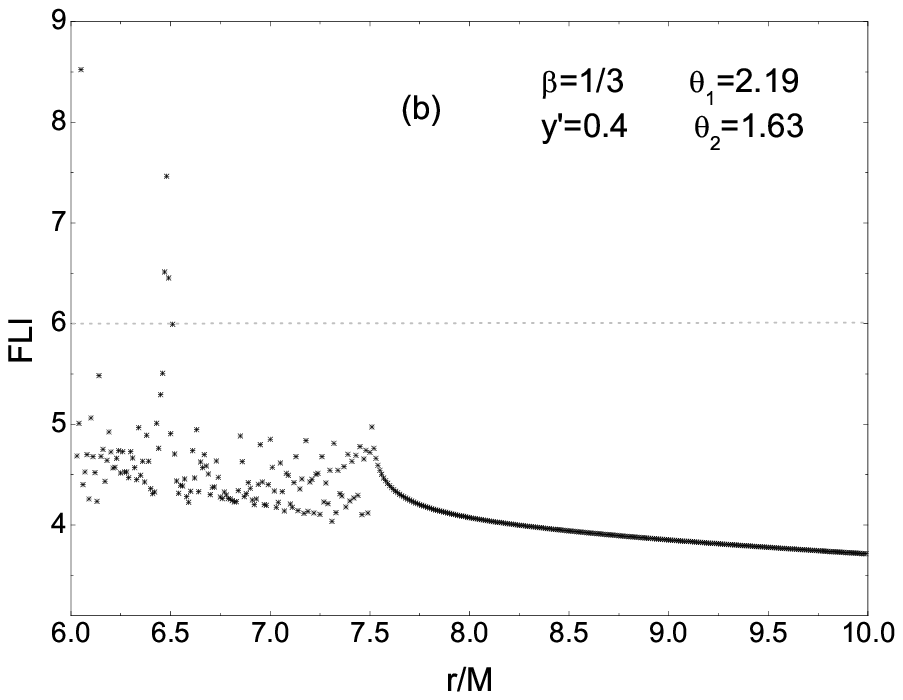}
\includegraphics[scale=0.75]{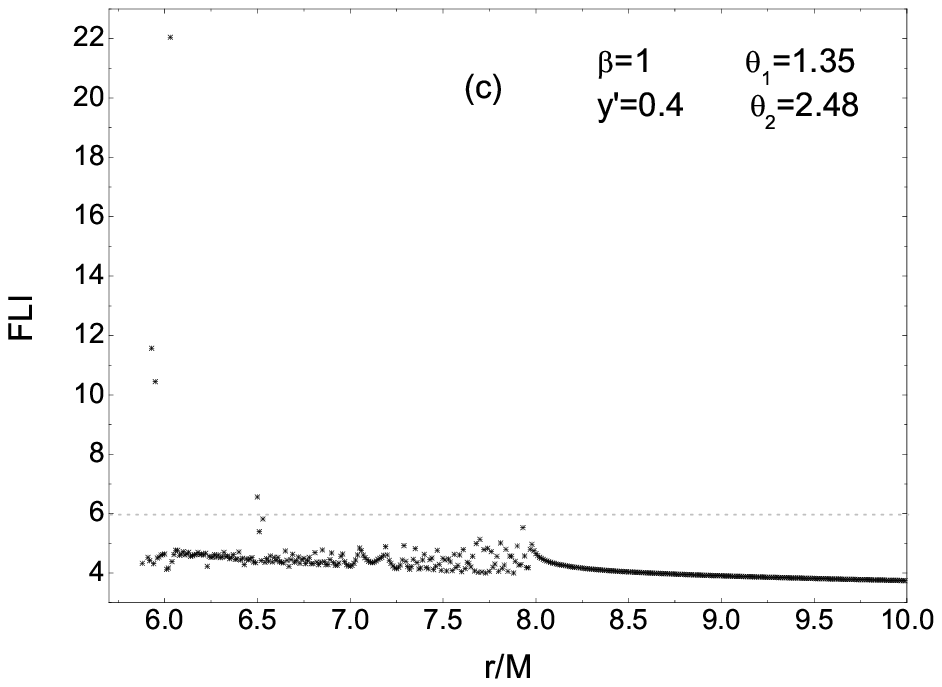}
\includegraphics[scale=0.75]{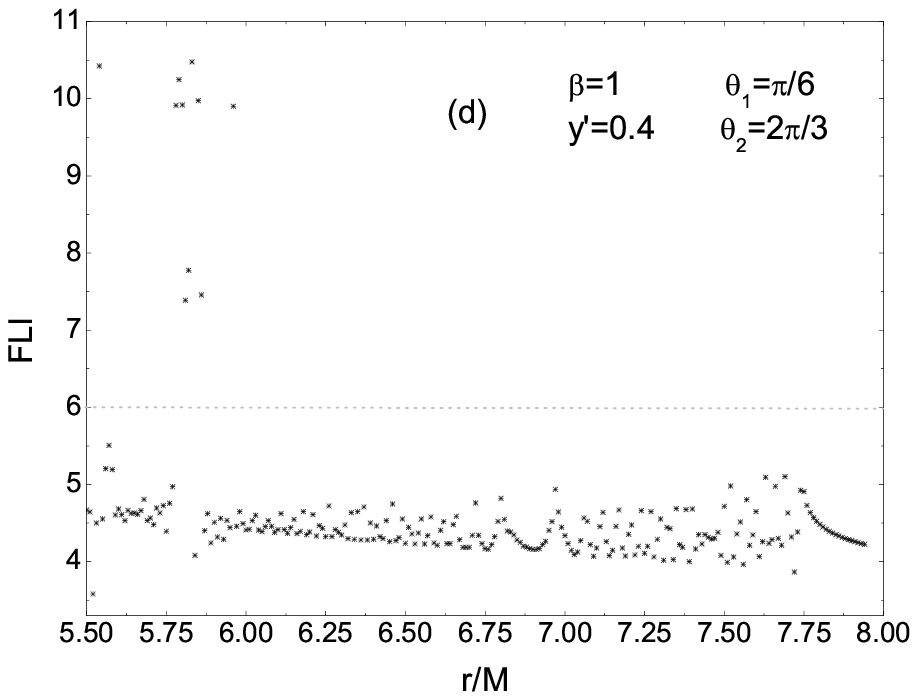}
\includegraphics[scale=0.75]{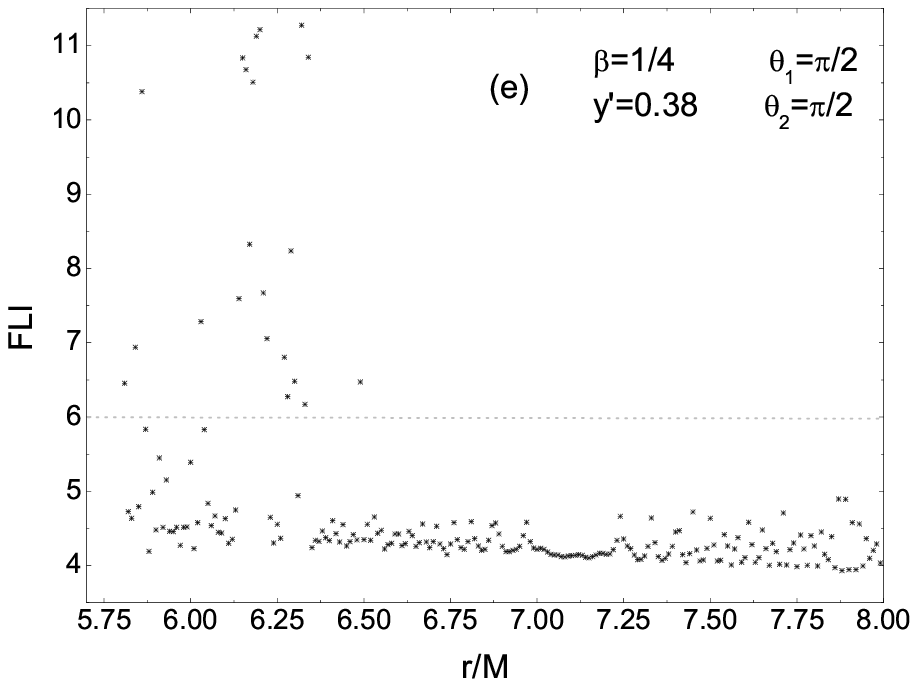}
\includegraphics[scale=0.75]{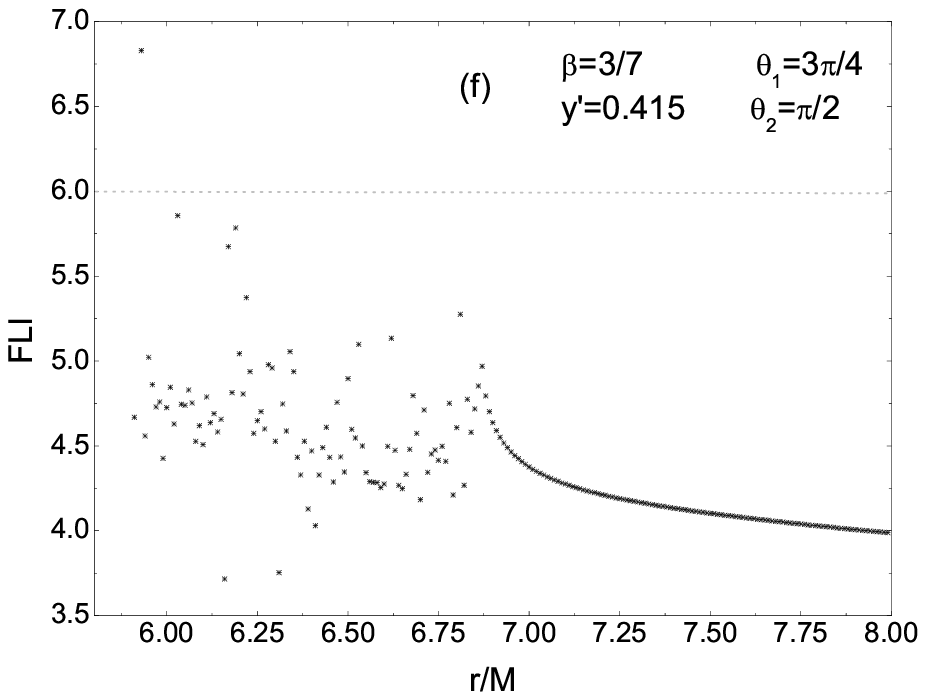}
\caption{FLI as a function of initial radius $r$ for various
parameters and other initial conditions. } \label{fig2}
\end{figure*}

\begin{figure*}[h]
\includegraphics[scale=0.75]{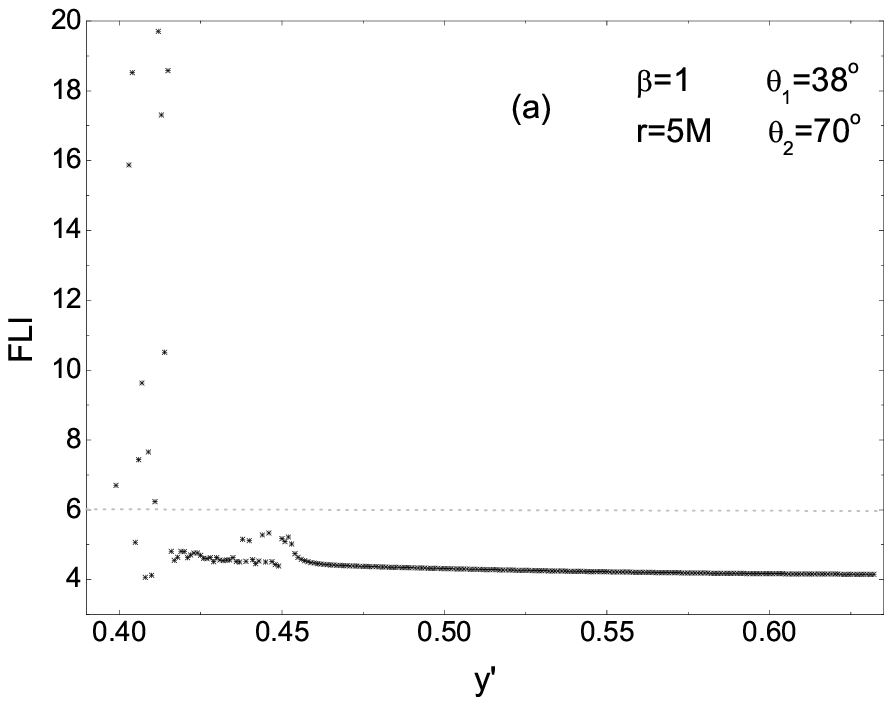}% Here is how to import EPS art
\includegraphics[scale=0.75]{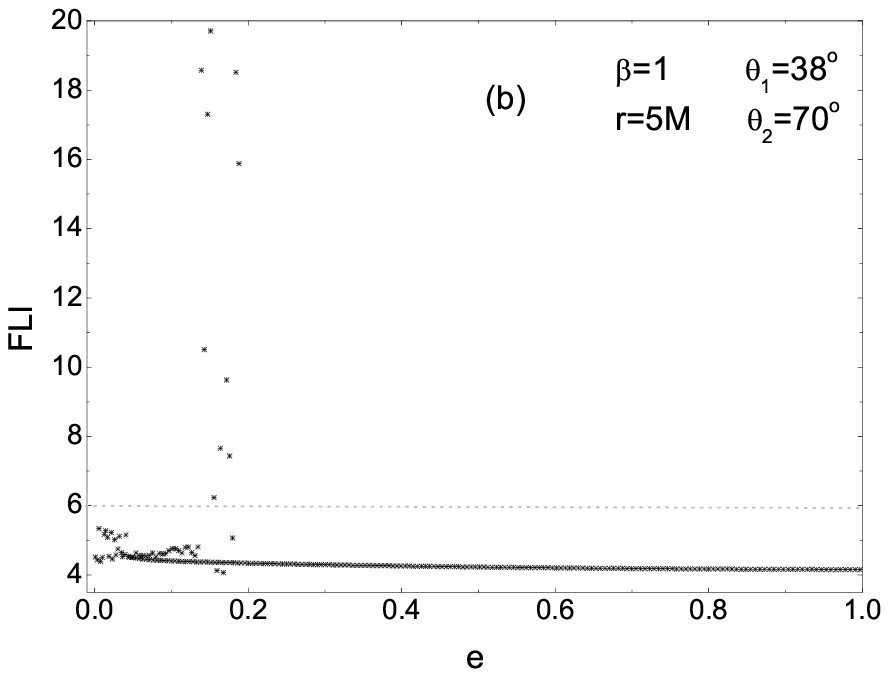}
\includegraphics[scale=0.75]{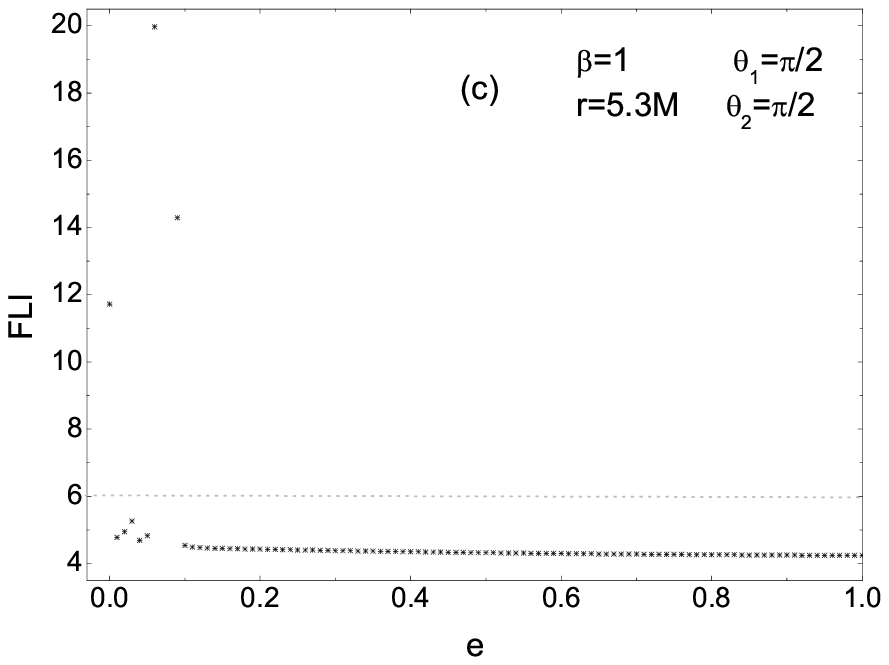}
\includegraphics[scale=0.75]{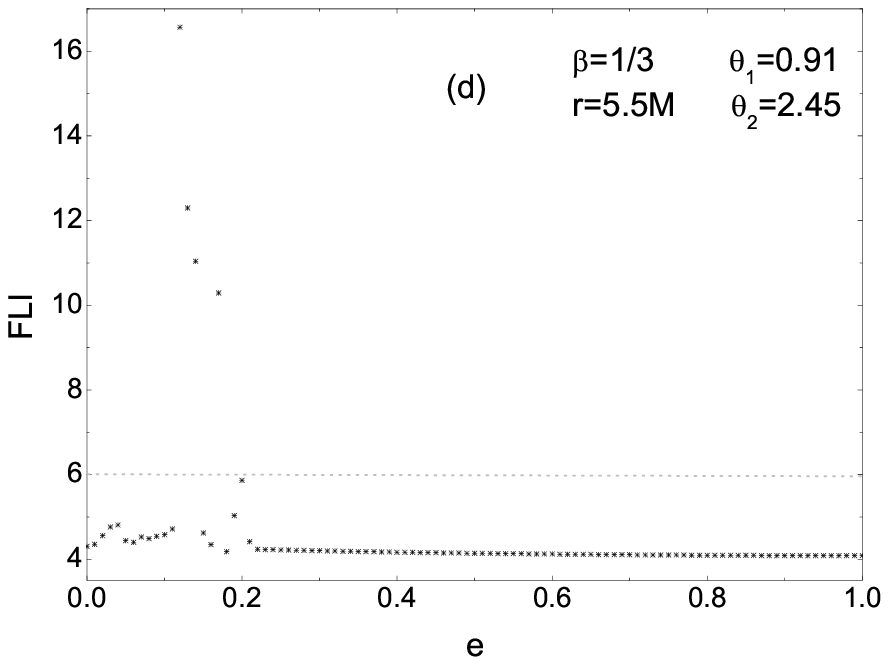}
\caption{(a) FLI as a function of initial velocity $\dot{y}$. (b)
same as (a) but for initial eccentricity $e$ in place of
$\dot{y}$. (c) and (d) relate to the cases of distinct parameters
and initial conditions adopted.} \label{fig3}
\end{figure*}

\begin{figure*}[h]
\includegraphics[scale=0.8]{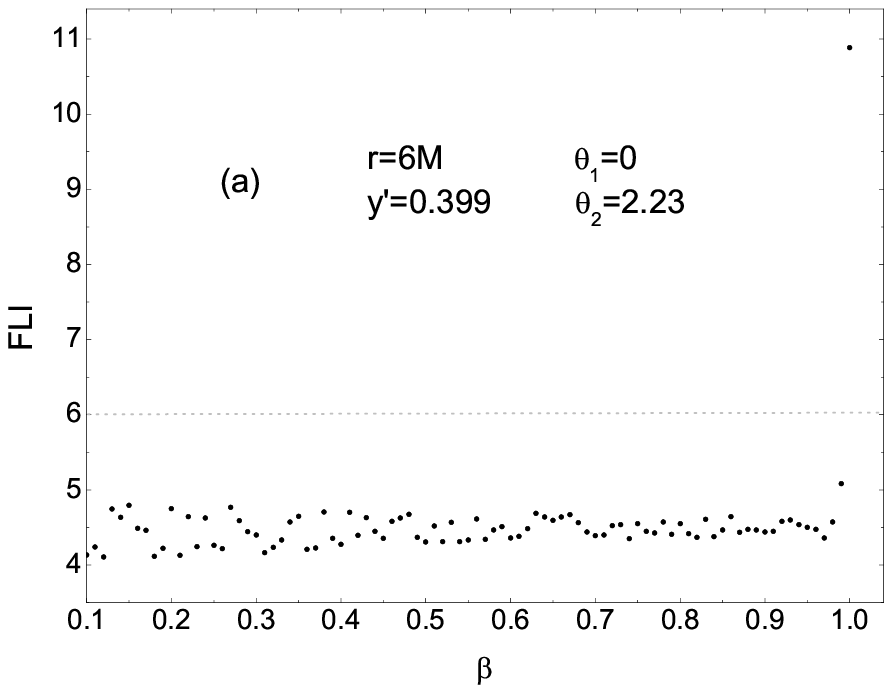}% Here is how to import EPS art
\includegraphics[scale=0.8]{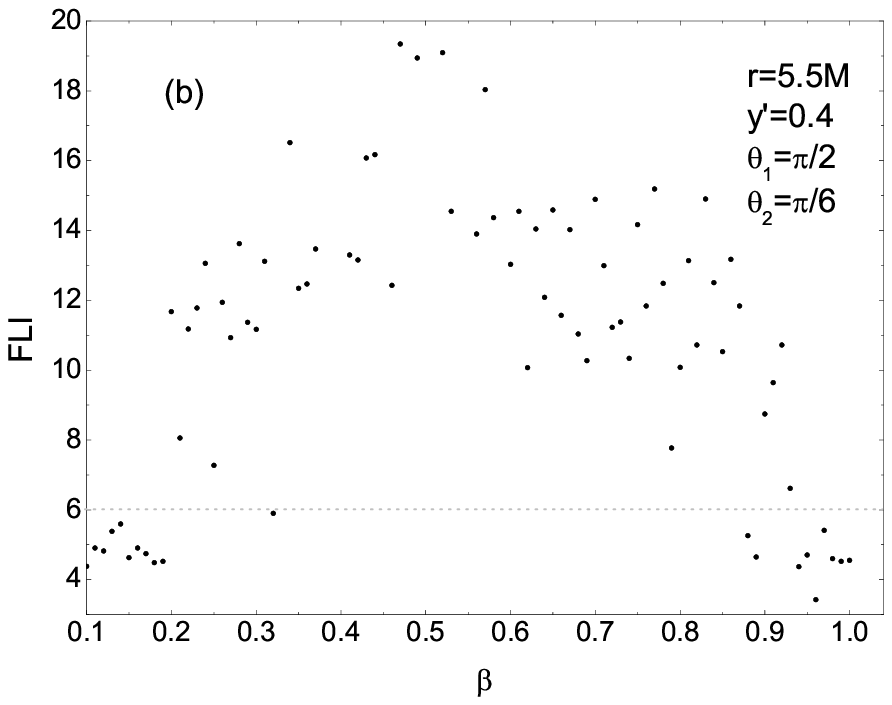}
\includegraphics[scale=0.8]{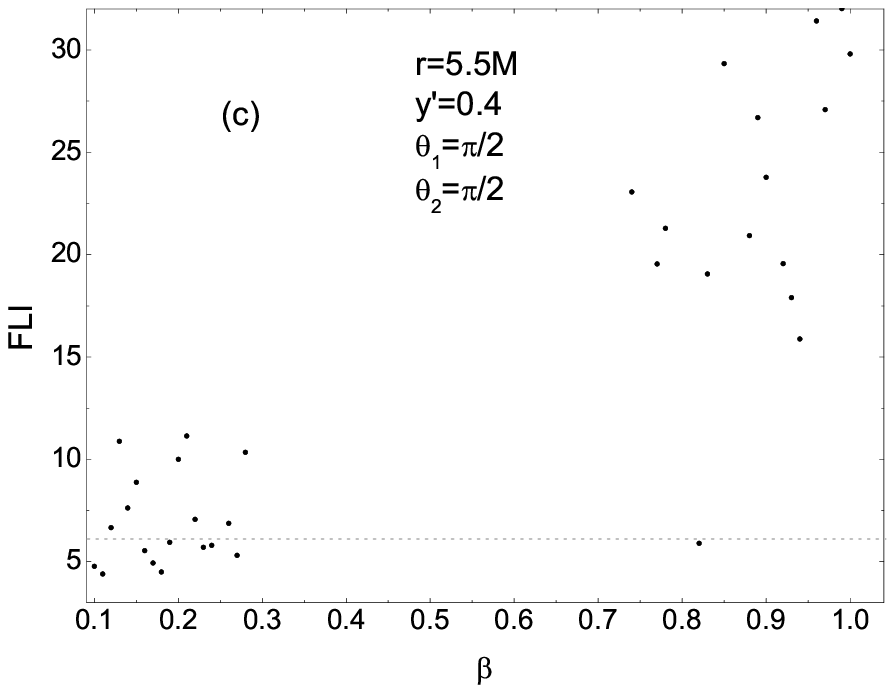}
\includegraphics[scale=0.8]{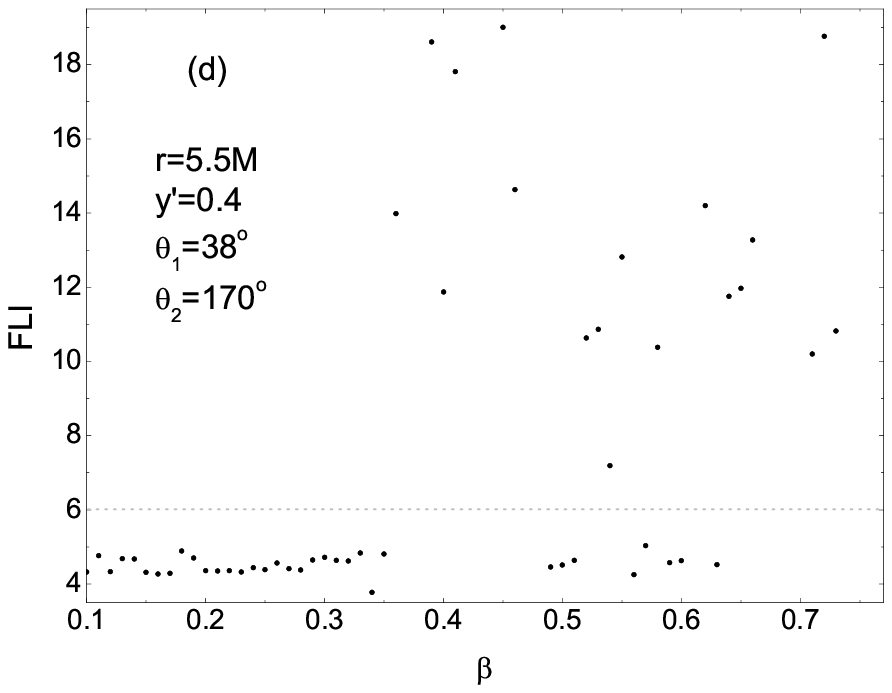}
\caption{FLI as a function of binary mass ratio $\beta$.
Specifically for panel (c), the blank interval with the binary
mass ratios in the range of $\sim (0.3-0.7)$ represents the
unstable coalescing orbits within the time considered. Some
details of the FLI about the unstable orbits are neglected.}
\label{fig4}
\end{figure*}

\begin{figure*}[h]
\includegraphics[scale=0.75]{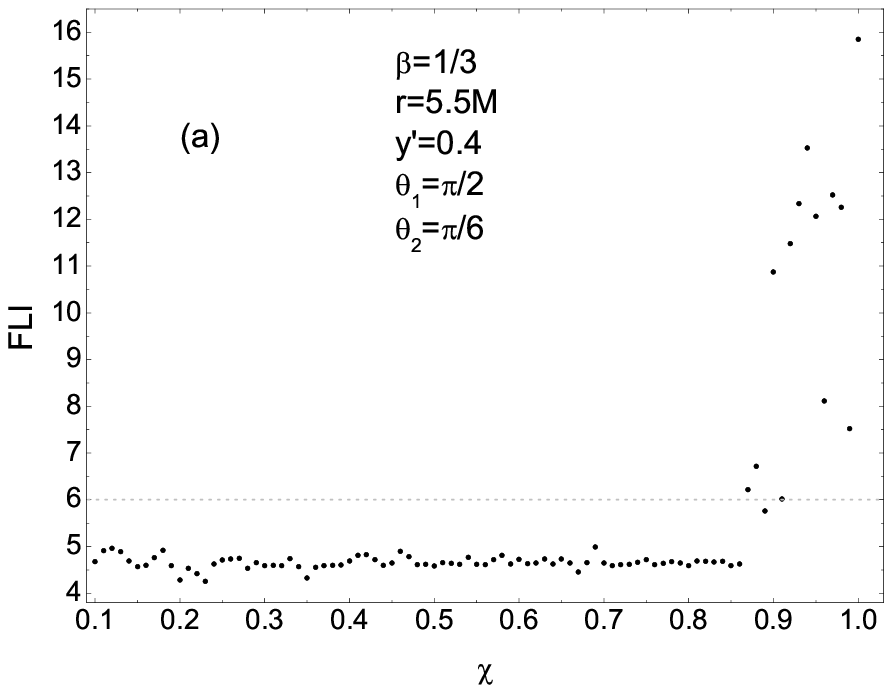}% Here is how to import EPS art
\includegraphics[scale=0.75]{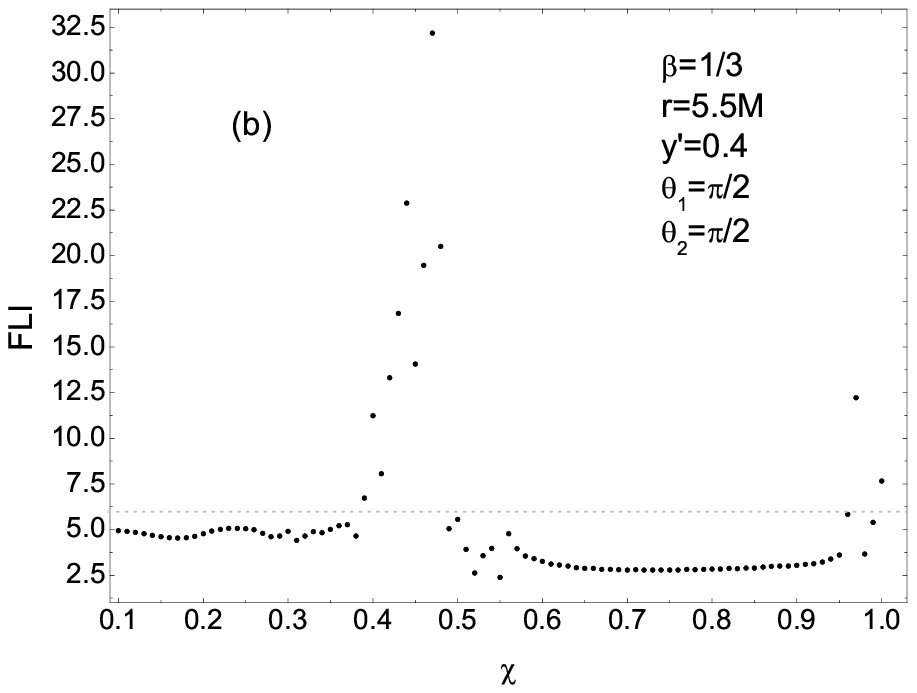}
\includegraphics[scale=0.75]{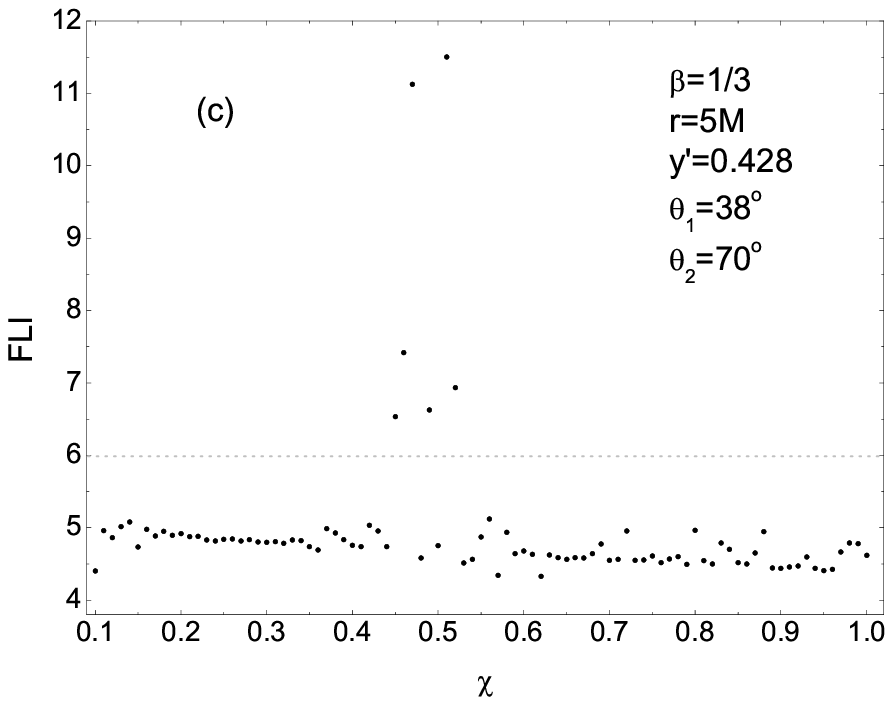}
\includegraphics[scale=0.75]{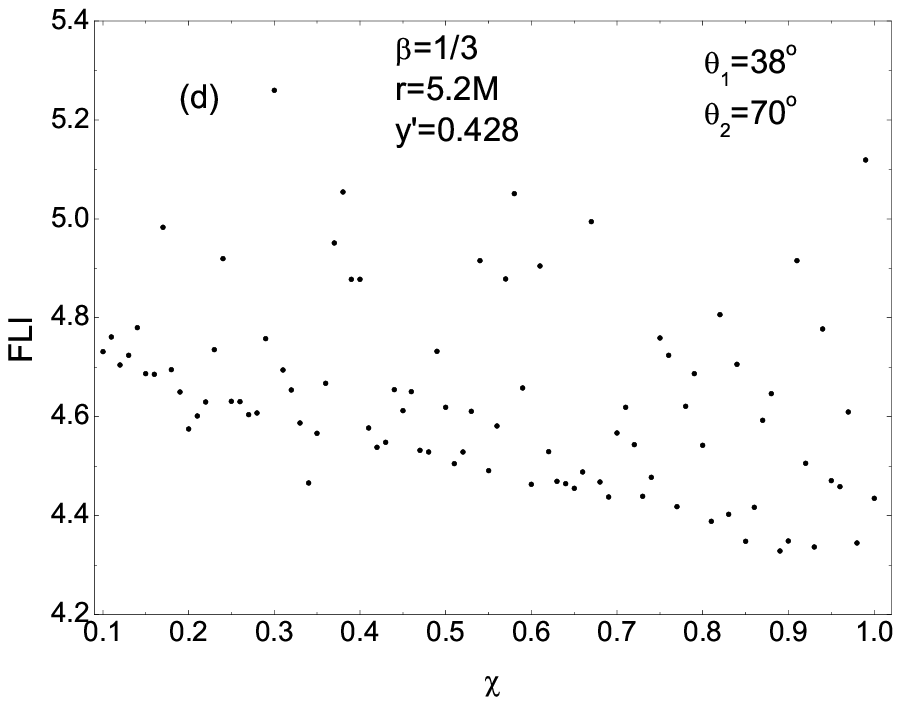}
\includegraphics[scale=0.75]{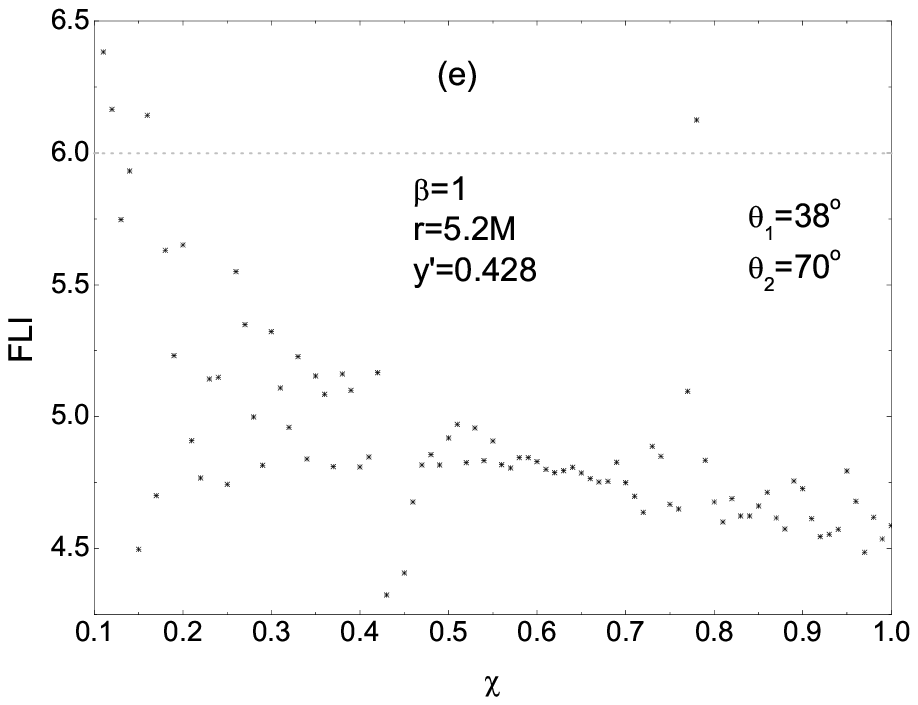}
\includegraphics[scale=0.75]{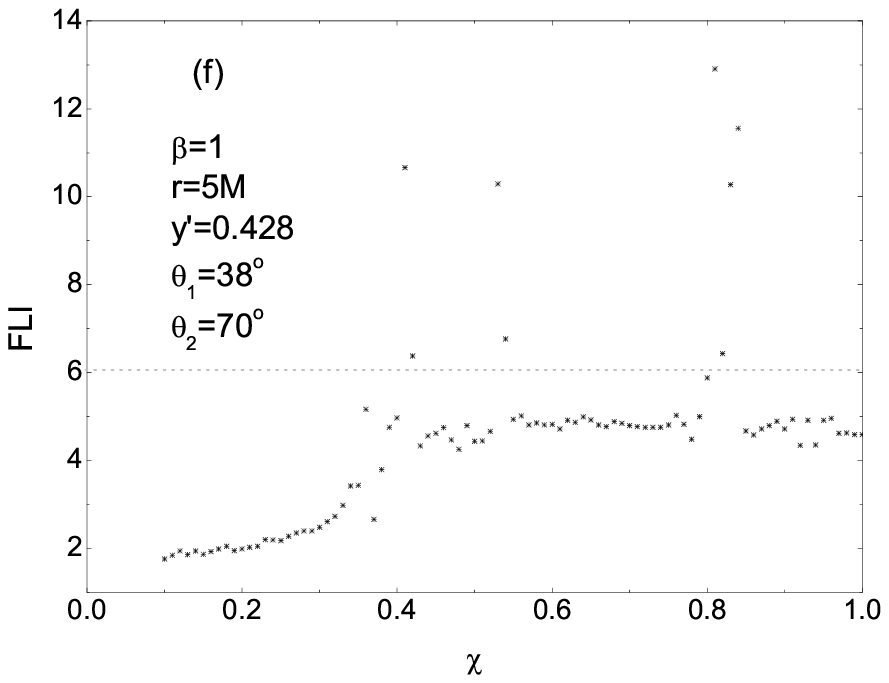}
\caption{FLI as a function of dimensionless spin parameters
$\chi=\chi_1=\chi_2$.} \label{fig5}
\end{figure*}

\begin{figure*}[h]
\includegraphics[scale=0.75]{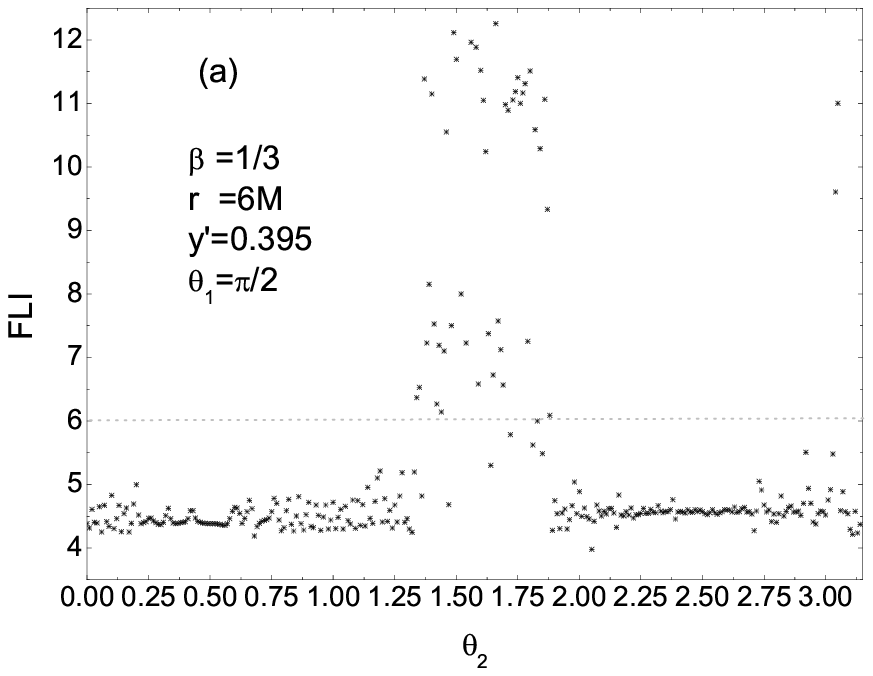}% Here is how to import EPS art
\includegraphics[scale=0.75]{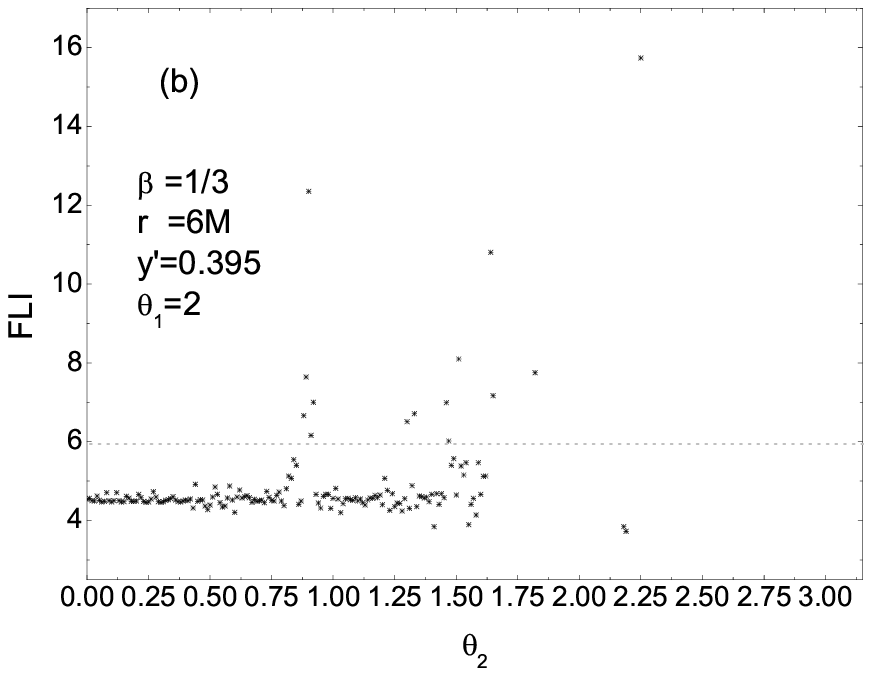}
\includegraphics[scale=0.75]{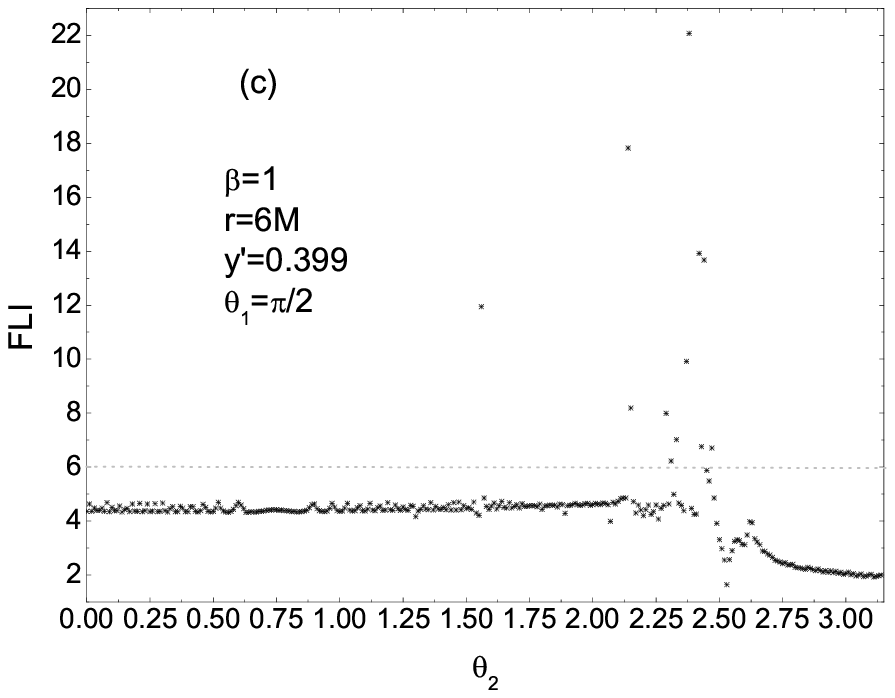}
\includegraphics[scale=0.75]{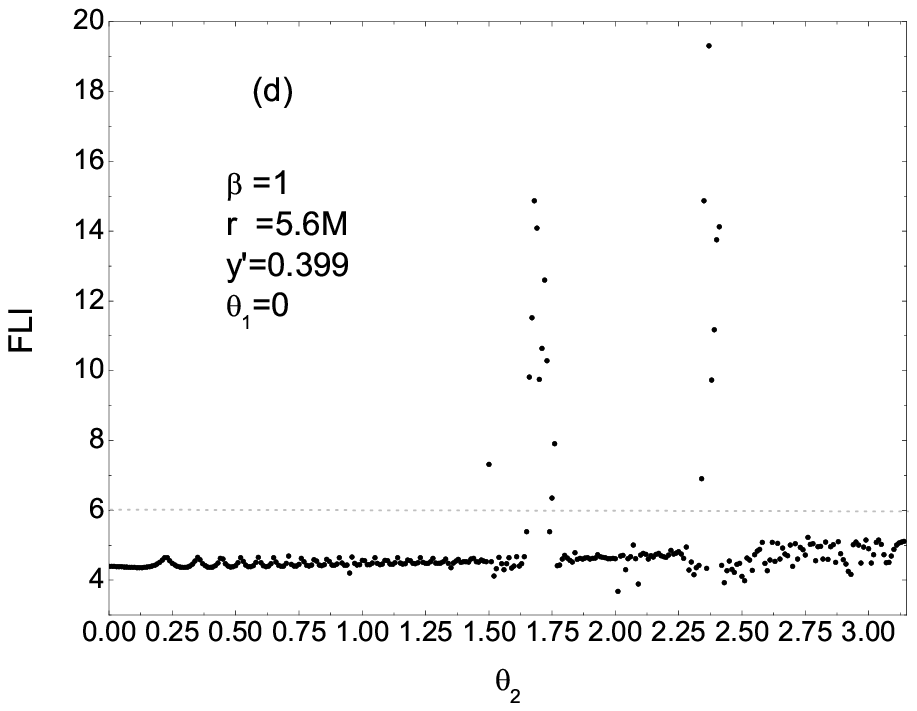}
\caption{FLI as a function of initial spin alignment $\theta_2$. }
\label{fig6}
\end{figure*}

\begin{figure*}[h]
\includegraphics[scale=0.75]{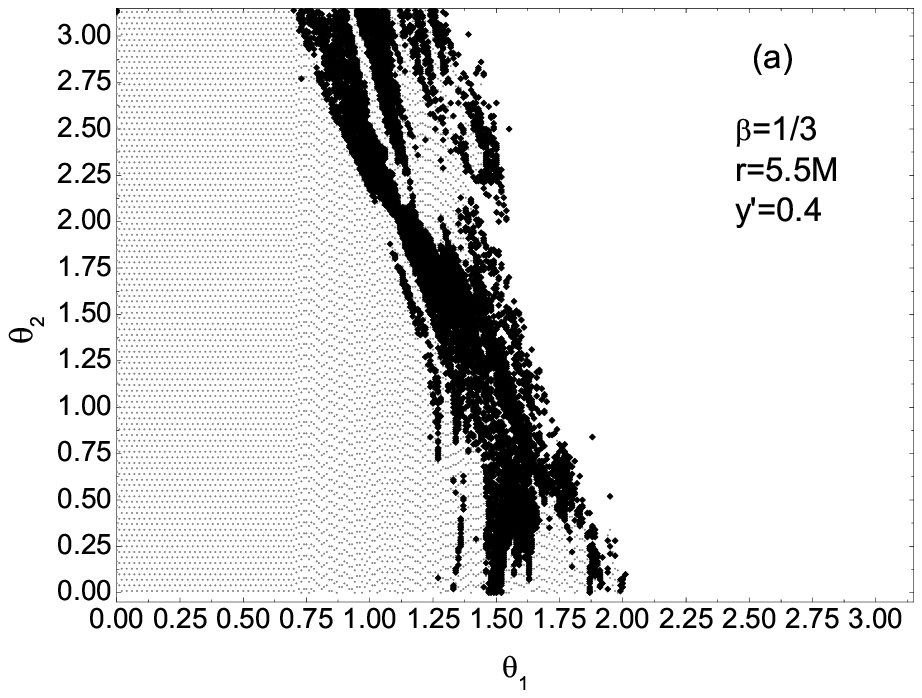}
\includegraphics[scale=0.75]{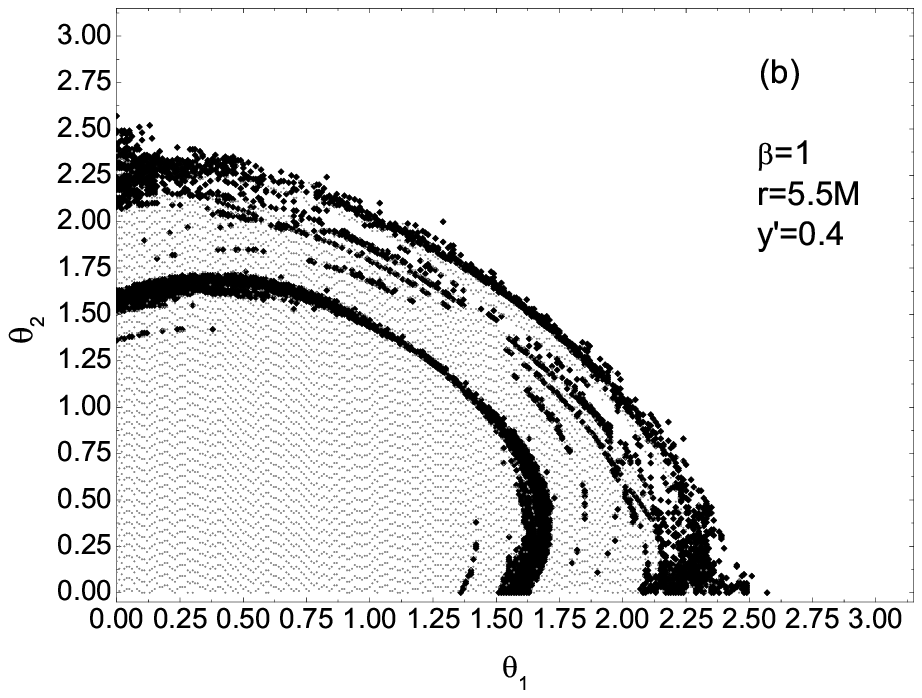}
\includegraphics[scale=0.75]{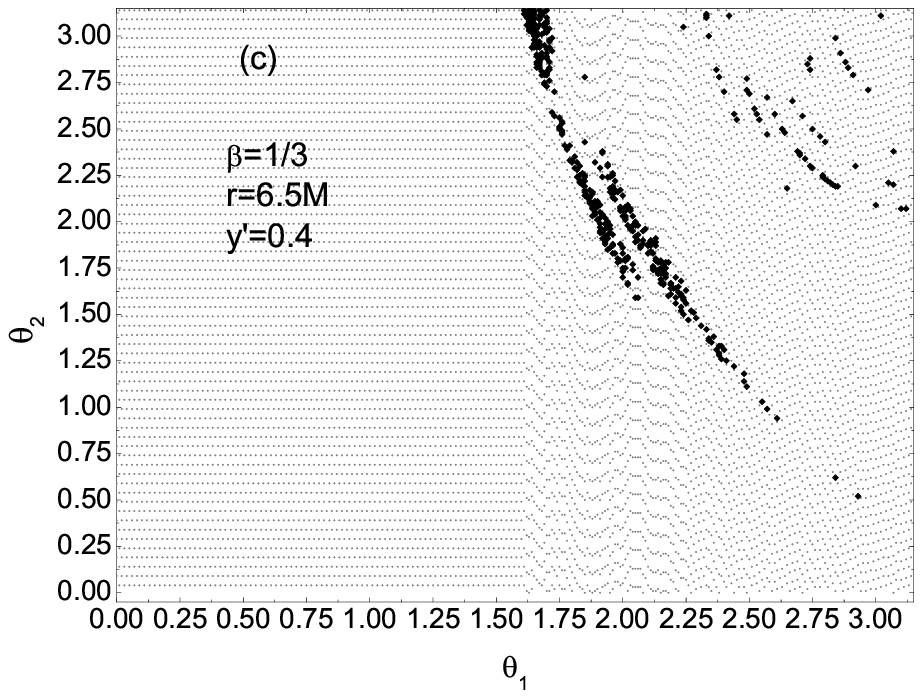}
\includegraphics[scale=0.75]{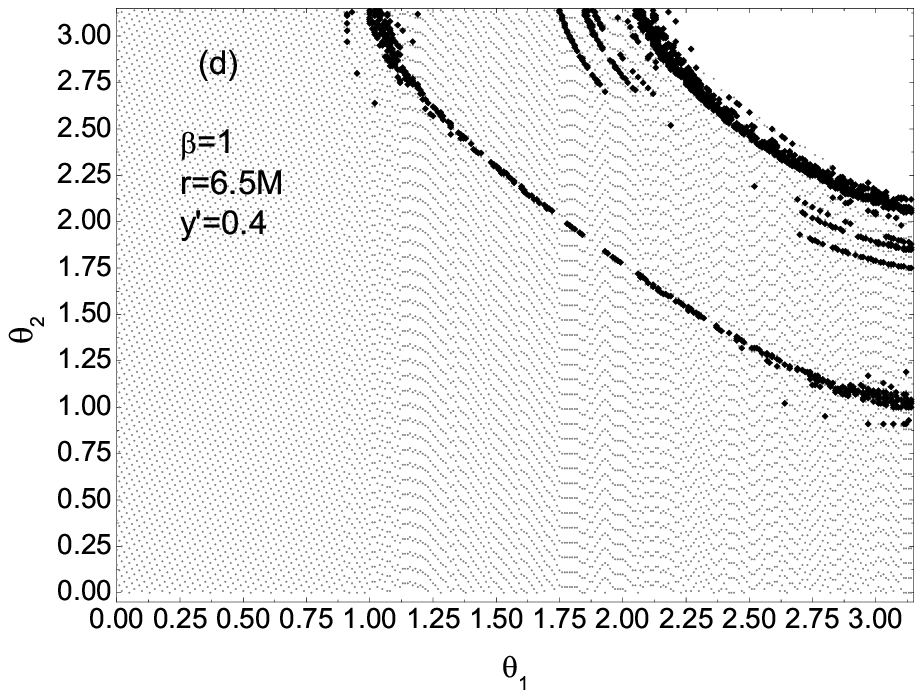}
\caption{Scans of many groups of initial points in the $(\theta_1,
\theta_2)$ plane. The black with $FLIs>6$ indicates chaos, but the
gray with $FLIs\leq 6$ signals regular. As an illustration,
$314\times314$ orbits are computed in each of panels, and initial
conditions color-coded white correspond to unstable merging pairs
during the time scale integrated. Of course, the color-coded white
regions should contain chaotic orbits as well as ordered ones. In
particular, all the orbits in panel (c) are stable. This implies
that the fractal basin boundary method becomes useless.}
\label{fig7}
\end{figure*}

\end{document}